\documentclass[review,3p,times,12pt]{elsarticle}
\usepackage{hyperref}
\usepackage{amsmath}
\usepackage{amsfonts}
\usepackage{graphicx,epsfig}
\usepackage{graphicx,subfigure}
\usepackage{natbib}
\usepackage{indentfirst}
\hypersetup{
    colorlinks = true,
               linkcolor = black,
            urlcolor  = black,
            citecolor = black,
            anchorcolor = black}
            \usepackage{epstopdf}
\usepackage[font=small,labelfont=bf,labelsep=space]{caption}

 \makeatletter
\input epsf
\newcounter{saveeqn}

\begin{document}
		\begin{frontmatter}
			
			\title{Flow dynamics in a wavy channel filled with anisotropic porous material under the effect of wall slip}
			
\author[add2]{Shyamal Kumar Mondal}
          \ead{shyamaliitm1989@gmail.com}
         \author[add3]{Sougata Mandal}
           \ead{mandalsougata5@gmail.com}
            \author[add1]{Sukhendu Ghosh \corref{cor1}}
    \ead{sukhendu.math@gmail.com}
			
			
       \address[add2]{Department of Applied Mathematics, University of Calcutta, Kolkata, West Bengal - 700009, India}
        \address[add3]{Department of Mathematics, Adani University, Ahmehdabad, Gujarat-382421, India}
          \address[add1]{Department of Mathematics, Indian Institute of Technology Jodhpur, Rajasthan - 342037, India}
    

\begin{abstract}
{In this study, a theoretical and graphical analysis is conducted to examine the effects of wall-velocity slip, anisotropic ratio, and porosity parameter on a two-dimensional, viscous, laminar, and incompressible flow through a wavy channel filled with anisotropic porous media. The flow is assumed to be steady and symmetric, with a constant volumetric flow rate imposed along the channel walls. The governing equations are described using the Darcy–Brinkman model coupled with the continuity equation, while the tangential velocity at the wavy boundaries is represented through Navier slip conditions.
An analytical solution is obtained using a perturbation approach under physically consistent boundary conditions. The effects of key parameters, including anisotropic ratio ($\lambda$), Darcy number ($Da$), and slip parameter ($\beta$), on flow characteristics such as axial velocity, pressure gradient, shear stress, and streamline patterns are examined in detail and presented graphically. The results indicate that wall velocity slip significantly reduces flow reversal, enhances near-wall velocity, and decreases the centerline velocity. For a fixed non-zero slip, a decrease in the Darcy number leads to a pronounced modification in the velocity profile, while increased slip further strengthens near-wall flow and weakens the core flow.
Additionally, the streamline analysis reveals that velocity slip plays an important role in controlling flow separation near the crest of the wavy wall. In the case of isotropic porous media with large amplitude wavy channel, flow separation can also be effectively regulated. Overall, the study demonstrates that velocity slip provides a powerful mechanism for controlling flow behaviour by altering the shear distribution within the perturbed flow, with potential applications in technological, geophysical, and biophysical transport systems.}

 \end{abstract}
			
			
\end{frontmatter}
		
\section{Introduction}
 The study of fluid flow through porous media has attracted considerable attention due to its wide range of applications in engineering, geophysical, industrial, and biological systems.
 It has various practical applications in filtration, thermal insulation, groundwater movement, oil recovery, and heat exchangers, etc.\cite{nield2006convection, gray2013darcy, datta2010study}. Porous media consist of a solid framework with interconnected voids and are commonly found in natural materials such as beach sand, sandstone, limestone, wood, and human lungs \cite{nield2006convection, huitt1956fluid, ng2010darcy}. The complexity of flow through porous media arises from fluid interactions with packing particles and boundary walls \cite{grattoni1995anisotropy}.Owing to its wide range of applications and the intricate physical behavior involved, porous media flow has attracted significant attention from researchers across diverse fields, including physics, mathematics, engineering, and other scientific disciplines. In this context, fluid flow in wavy channels has also been extensively investigated under various physical conditions \cite{tsangaris1984laminar, kitanidis1997stokes, malevich2006stokes}, as such studies provide valuable insights into the dynamics and transport characteristics of complex flow systems. Wang et al. \cite{yu2013darcy} investigated forced flow through a bumpy-walled channel containing a porous medium. Their model is relevant to microfluidic systems where surface roughness becomes significant at small scales. They applied a perturbation approach to solve the Darcy-Brinkman equations, using the ratio of bump amplitude to channel width as a key parameter. Their analysis examined the effect of the non-dimensional porous medium parameter on the flow dynamics. The solutions represent the clear fluid limit as the non-dimensional porous medium parameter approaches zero and the Darcy limit as it approaches infinity. In the context of wavy channels filled with isotropic porous materials, research has also focused on hydrodynamics and heat transfer, providing deeper insights into fluid behavior in such systems \cite{kumar2020elastohydrodynamics, karmakar2021physics}. The real-world porous media often exhibit anisotropic permeability, where fluid flow differs in horizontal and vertical directions due to grain orientation \cite{rice1970anisotropic}. The anisotropic permeability ratio $\lambda$, defined as the ratio of horizontal to vertical permeability, has been extensively investigated in many studies \cite{karmakar2021physics, rice1970anisotropic, yovogan2013effect} to understand its impact on flow behavior. The studies by Ng and Wang \cite{ng2010darcy} and Karmakar et al. \cite{karmakar2016effect} have demonstrated that the anisotropic permeability ratio $\lambda$ plays a crucial role in influencing flow dynamics, especially in anisotropic porous layers. Additionally, numerous studies have highlighted the significant influence of porosity on flow dynamics, making it a key area of research.
Haddad et al. \cite{haddad2007forced} numerically investigated the steady laminar forced convection slip-flow of an incompressible Newtonian gas through a porous microchannel under Local Thermal Non-Equilibrium (LTNE) conditions. They demonstrated the effects of key parameters, including Knudsen number, porosity of the media, Darcy number, Forchheimer number, Peclet number, Biot number, and thermal conductivity ratio, on velocity slip, temperature jump, skin friction, and heat transfer. Moreover, The study of Khurram et al.\cite{javid2022porosity} examined the steady, laminar peristaltic flow of a non-Newtonian biological fluid in a porous channel, and emphasized the effects of porosity and couple stress parameters in the flow behavior. Their analytical solutions reveal that increased porosity enhances velocity and peristaltic pumping, while the couple stress parameter reduces viscous effects. Additionally, Degan et al. \cite{degan2002forced} analyzed how porosity influences velocity, pressure gradients, and heat transfer in anisotropic porous channels. Further studies exploring the effects of porosity are available in \cite{karmakar2016effect, tripathi2013study, zeeshan2018convective}.

  
In earlier discussions, most analyses, whether analytical, theoretical, or experimental, have primarily assumed no-slip conditions at the channel walls. However, much less attention has been given to cases involving slip boundary conditions, which introduce additional complexities to the considered problems. Wall slip is highly relevant in various real-world applications, such as lubrication, microfluidics, drag reduction on biological and technological surfaces, high-speed rarefied flows, and polymer melts. In these cases, a viscous fluid may exhibit tangential slip along the wall, leading to significant deviations from classical no-slip models. Slip boundary conditions become particularly important when fluid-solid interactions occur at small scales or when there is relative motion between the fluid and the solid wall surface. This type of boundary condition is commonly used in microfluidic porous media systems and various biological flow systems, such as arterial blood flow systems. In such scenarios, the conventional no-slip assumption may not hold, necessitating the use of slip models to describe the flow behavior accurately \cite{zhu2002limits, sharipov2004velocity, bonaccurso2002hydrodynamic, pit2000direct}.
The extent of slip at the interface is largely determined by its structural and dynamic characteristics, which play a crucial role in shaping the overall flow behavior. In 1824, Navier \cite{navier1822memoire} proposed a slip boundary model in which slip velocity is directly proportional to shear stress at the wall, expressed as $U_x = \beta \frac{\partial V}{\partial y}$,
 where the coefficient \(\beta\), known as the slip length, represents the extrapolated distance at which fluid tangential velocity becomes zero. The no-slip condition holds when $\beta=0$, whereas $\beta \neq 0$ indicates the presence of slip at the boundary. However,  Slip effects are most prominent on superhydrophobic surfaces, inflows of non-Newtonian fluids, hydrophobic surfaces with trapped nanobubbles, and under high shear rates, but they are less pronounced on hydrophilic, rough surfaces. Pitrowski \cite{helmholtz1860reibung} introduced a coefficient to quantify slip near walls, while Rao and Rajagopal \cite{rao1999effect} examined the influence of shear and normal stresses in modeling slip within channels. Migler et al. \cite {migler1993slip, migler1994slip} measured slip at interfaces using evanescent wave fluorescence. Furthermore, Prakash and Raja Sekhar \cite{kumar2020elastohydrodynamics} studied unidirectional elastic-hydrodynamics in deformable porous media, analyzing the effects of induced shear and pressure gradients. Min and Kim \cite{min2005effects} applied slip velocity at channel walls to analyze stability in two-fluid plane Poiseuille flow, while Nield and Bejan \cite{nield2006convection} provide a comprehensive literature survey on the topic. Usha and Chattopadhyay \cite{chattopadhyay2016yih} demonstrated the potential to control interfacial instabilities by designing channel walls with slip properties, and Ghosh et al. \cite{ghosh2014linear} showed that hydrophobic wall surfaces can stabilize or destabilize flows when modeled with velocity slip at the walls.

Notably, most of the existing studies focus on fluid flow through various channels but rarely consider the combined effects of velocity slip, porosity, anisotropy, and wavy channel geometry on flow dynamics. Motivated by the above literature survey, we extend the findings of Karmakar et al. \cite{karmakar2017note} by integrating slip conditions at the wall.
The main goal of this study is to explore how the flow dynamics within a wavy channel are influenced by hydrodynamic parameters, specifically the anisotropic ratio \(\lambda\), slip parameter \(\beta\), and porosity parameter \(Da\). We examine how these parameters affect axial velocity, wall pressure gradient, shear stress distribution, and streamline variations. The problem is first non-dimensionalized, followed by a perturbation analysis for small values of \(\delta\), where \(\delta\) represents the ratio of the channel’s mean width \(b\) to its wavelength \(L\). Throughout this study, we investigate the effects of velocity slip, porosity, anisotropy, and channel geometry on flow dynamics, offering deeper insights into the underlying physical mechanisms. \\
{The paper is arranged categorically as follows: After the introduction, the problem is defined and the mathematical formulation of the physical problem is provided in the appropriate coordinate system with the accompanying boundary conditions. In Section \ref{sec3}, the perturbation technique is used to solve the system. The flow behaviours away from the wavy wall are illustrated in Section \ref{sec4}. Section \ref{sec5} centers on the discussion of the analytical results graphically. This part involves validation using accessible literature results. The Reverse flow region is presented in Section \ref{sec6}. Some limiting cases are discussed in Section \ref{sec7}, while the summary and findings are concluded in Section \ref{sec8}. }



\section{Mathematical Formulation}\label{sec2}
		
\subsection{Governing equations}
\begin{figure}[h] 
\centering
\includegraphics[height = 6.5cm]{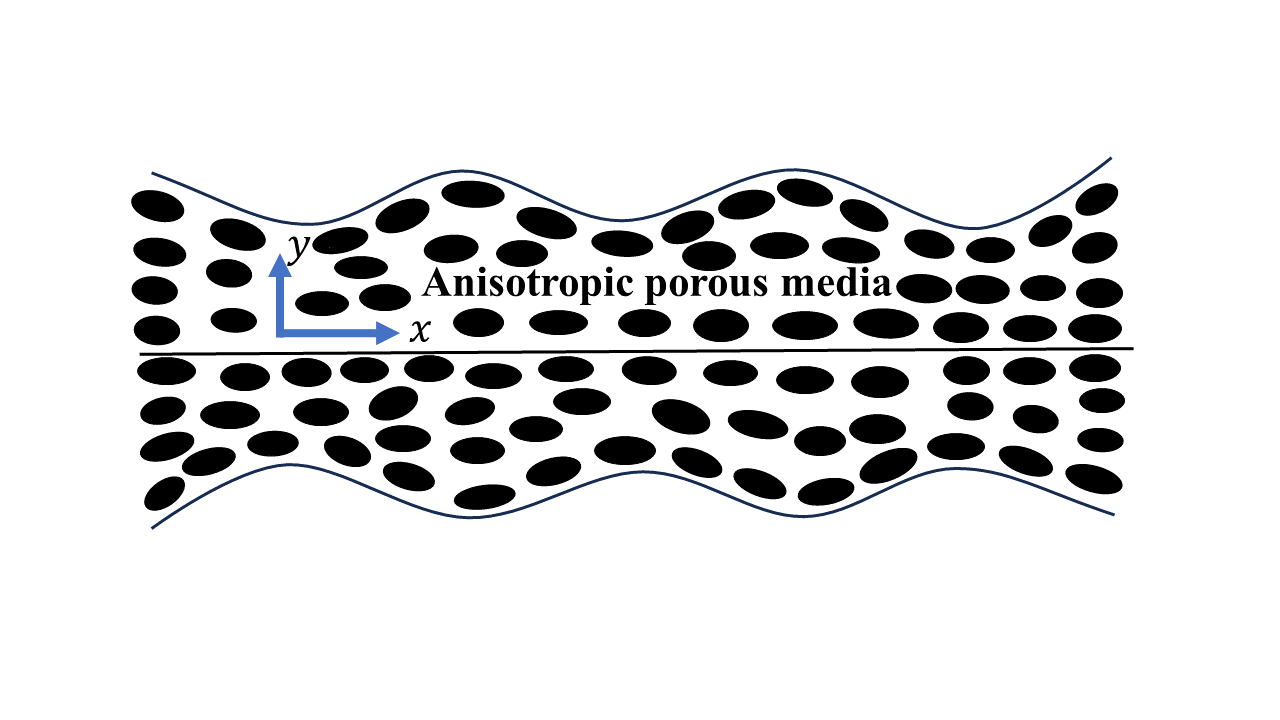}
\caption{Physical configuration of the wavy-walled channel filled with anisotropic porous material.}
\label{Threshold_fig 111}
\end{figure}

In this section, we consider a two-dimensional flow of a Newtonian, incompressible fluid in a wavy-walled channel filled with anisotropic porous material, incorporating wall slip effects. The flow is symmetric about the centerline of the channel; therefore, the mathematical formulation is developed only for the upper half of the domain of the wavy channel (see Fig.~\ref{Threshold_fig 111}). A Cartesian coordinate system is adopted by considering $x^*$ and $y^*$ directions along and perpendicular to the centerline of the channel ($y^* = 0$). The boundaries of the channel are two impermeable walls located at $y^* = h(x^*) = b(1+a\cos(2 \pi x^*/L))$ and $y^* = -h(x^*) = -b(1+a\cos(2 \pi x^*/L))$, respectively. Here, $b$ represents the channel's mean width. The channel wall's amplitude is $b a$, and its wavelength is $L$. Owing to the above considerations, the equations governing the considered steady flow of the viscous fluid are given by the Darcy-Brinkman momentum equations and the equation of the conservation of mass \cite{karmakar2016effect,  degan2002forced, karmakar2017note, karmakar2016lifting} as follows: 
\begin{eqnarray}\label{e1}
\nabla^*p^*-\mu_{eff}\nabla^2 \textbf{u}^*+\mu \textbf{K}^{-1}\textbf{u}^* = 0, 
\end{eqnarray}
 \begin{eqnarray}
 \nabla^*.\textbf{u}^* = 0,	
\end{eqnarray}
where $p^*$ and $\textbf{u}^*$are dimensional pressure and velocity vectors respectively, $\mu$ is the dynamic viscosity of the fluid and $\mu_{eff}$ is the effective viscosity of the fluid in the porous media. In the present study, to prevent confusion and concentrate mainly on the impacts of velocity slip $\beta$, anisotropic ratio $\lambda$, and porosity parameter $Da$ within the corrugation, we consider the assumption $\mu$ = $\mu_{eff}$ \cite{nield2006convection}. In addition, \textbf{K} represents the symmetrical second-order permeability tensor \cite{karmakar2016effect, karmakar2017note, karmakar2016lifting} of the porous matrix provided by:
		
\begin{eqnarray}
\textbf{K} = \begin{pmatrix}
K_{1} & 0\\
0 & K_{2}\\
\end{pmatrix},
\end{eqnarray}
where $K_{1}$ and $K_{2}$ represent the permeabilities along horizontal and vertical axes, respectively. Notably, permeability is scalar quantity in an isotropic porous media \cite{nield2006convection}.

Let $Q$ be the characteristic volumetric flow rate per unit depth into the channel, which is given by 
       \begin{eqnarray}
		Q = \int_{0}^{h(x^*)}u^*dy^*.
		\end{eqnarray}
\subsection{Boundary conditions}
Since the flow system is symmetric around $x^*$-axis ($y^* = 0$), the considered problem is analyzed in the half-domain approach ($0 \le y \le h(x^*)$). Consequently, the Navier slip boundary conditions are applied to the wavy wall  $y^* = h(x^*) = b(1+a\cos(2\pi x^*/L))$ only, which are defined as follows: 
\begin{eqnarray}
u^* = \beta^*{u^*_{y^*}}, ~v^* = 0. 
\end{eqnarray}
Here, $\beta^*$ is the dimensional slip parameter, often known as the slip length or slip coefficient, that depends on the structure and characteristics of the surface. When $\beta^* = 0$, the widely assumed no-slip boundary conditions are applied on the surface. In this study, we assume that fluid slip occurs exclusively on the wavy surface within the pore spaces. Thus, The symmetry boundary conditions at $y^* = 0$ is defined by
\begin{eqnarray}
\frac{\partial u^*}{\partial y^*} = 0, ~v^* = 0.
\end{eqnarray}
\subsection{Non-dimensionalization}
Non-dimensionalization is usually employed to simplify the analytical complexity of visualizing the considered Darcy–Brinkman model. Before proceeding with the solution, the problem is reformulated in terms of dimensionless variables and parameters, defined as follows:
$$x = \frac{x^{*}}{L},  ~y = \frac{y^{*}}{b},  ~\delta = \frac{b}{L}, ~p = \frac{p^*}{\frac{\mu QL}{K_1b}},$$
$$  (u, v) = \left(\frac{u^{*}}{\frac{Q}{b}}, ~\frac{v^{*}}{\frac{Q}{L}}\right), ~\beta = \frac{\beta^*}{b}, ~Da = \frac{K_{1}}{b^{2}} ,  ~\lambda^{2} = \frac{K_{1}}{K_{2}}.$$
Herein, $ \delta$ is the ratio of the channel's mean width to its wavelength which is referred to as the aspect ratio, $Da$ is the Darcy number which reflects how easily the flow may percolate and $\lambda^{2}$ is the ratio of the permeability along $x$-direction to the permeability along $y$-direction, which is considered the anisotropic ratio. In the present study, we pay attention to some key flow parameters: $\beta$, $Da$, and $ \lambda$, which characterize the velocity slip, Darcy number/porosity parameter, and anisotropic ratio, respectively.
		
The resultant dimensionless governing equations in component form and accompanying boundary conditions are (after suppressing $\ast$)
\begin{eqnarray}\label{e2}
\alpha^{2}u + \alpha^{2}\frac{\partial p}{\partial x} - \delta^{2}\frac{\partial^2 u}{\partial x^2} - \frac{\partial^2 u}{\partial y^2} = 0, 
\end{eqnarray}
\begin{eqnarray}\label{e20}
\delta^2\lambda^2\alpha^2v + \alpha^2\frac{\partial p}{\partial y} - \delta^4\frac{\partial^2v}{\partial x^2} - \delta^2\frac{\partial^2v}{\partial y^2} = 0,
\end{eqnarray}
\begin{eqnarray}\label{e21}
\frac{\partial u}{\partial x} + \frac{\partial v}{\partial y} = 0,    
\end{eqnarray}
where $\alpha^2 = \frac{1}{Da},$\\ 
and 
\begin{eqnarray}\label{e3}
u = \beta{u_y}, ~v = 0 ~\mbox{at}~ y = h(x) = 1+a\cos(2\pi x).
\end{eqnarray}
It is important to note that as $\beta$ $\rightarrow 0$, the condition corresponds to a typical no-slip boundary. In contrast, when $\beta$ $\rightarrow \infty$, the wavy boundary of the porous medium experiences no shear. 

Further, the non-dimensional symmetry boundary conditions are given by 
\begin{eqnarray}\label{e4}
\frac{\partial u}{\partial y} = 0, ~v = 0 ~\mbox{at}~ y=0,
\end{eqnarray}
and the dimensionless volumetric flow rate condition is chosen as 
\begin{eqnarray}\label{e5}
\int_{0}^{h(x)} u dy = 1 .
\end{eqnarray}
\section{Method of Solution}\label{sec3}
The problem can be solved by considering the boundary conditions in two ways: (i) by considering the entire domain between the two wavy walls or (ii) by utilizing the half-domain approach, leveraging the symmetry of the flow about the channel centerline. For simplification, the latter method is adopted, where boundary conditions are applied at the centerline $y=0$ and the wavy wall $y=h(x)$, instead of at both wavy walls $y=\pm h(x)$. This approach ensures that the analysis is confined to symmetric slip conditions, with an identical slip length $\beta$  for both walls.
We now focus on solving the flow system governed by the equations and boundary conditions defined in \eqref{e2}-\eqref{e4}. Notably, after transforming the governing equations into dimensionless form, these equations of motion are still in complex form. As a result, it is complicated for a mathematician to find the analytical solution directly. To overcome this challenge we shall use the perturbation theory for solving the considered flow model. Since the axial variation of the wavy wall is assumed to be small enough, i.e.  $\delta$ $\ll 1$, this permits us to apply the perturbation technique \cite{ng2010darcy, tsangaris1984laminar, karmakar2017note, wei2003flow, luo2008two} to our problem. Consequently, the flow quantities can be expanded in a perturbation series for small $\delta$ as follows: 
\begin{eqnarray} \label{e6}
(u,v,p) = (u_0, v_0, p_0) + \delta^2 (u_1, v_1, p_1) + O(\delta^4).
\end{eqnarray}
The following subsections are the outcomes of substituting \eqref{e6} into \eqref{e2} - \eqref{e5} and collecting the quantities of the same order in $\delta$.
\subsection{The leading order problem}
For the leading-order problem, the governing equations \eqref{e2} - \eqref{e21} becomes
\begin{eqnarray} \label{e7}
\alpha^2u_0  +  \alpha^2\frac{\partial p_0}{\partial x} - \frac{\partial^2u_0}{\partial y^2} = 0,
\end{eqnarray}
\begin{eqnarray}
-\frac{\partial p_0}{\partial y} = 0,
\end{eqnarray} 
\begin{eqnarray}\label{mcon1}
\frac{\partial u_0}{\partial x} + \frac{\partial v_0}{\partial y} = 0.
\end{eqnarray}
The corresponding leading order boundary conditions for \eqref{e3} and \eqref{e4} are given by:
\begin{eqnarray}\label{bdcn1}
u_0 = \beta u_{0y},  ~v_0 = 0,~\mbox{at}~ y = h(x) = 1+a\cos(2\pi x),
\end{eqnarray}
and, 
\begin{eqnarray}\label{bdcn2}
\frac{\partial u_0}{\partial y} = 0, ~v_0 = 0 ~\mbox{at}~ y=0.
\end{eqnarray}
The  leading order volumetric flow rate condition is given by
\begin{eqnarray}
\int_{0}^{h(x)}u_0(x,y)dy = 1.
\end{eqnarray}
With the help of the above boundary conditions, the leading order flow fields can be solved. The solution  of the equation \eqref{e7} for $u_0$ subject to the above boundary conditions \eqref{bdcn1}-\eqref{bdcn2} is given by
\begin{eqnarray} \label{e13}
u_0(x,y) = -p_{0x}+a_1(x)\cosh(\alpha y)+a_2(x)\sinh(\alpha y),
\end{eqnarray}
where
\begin{eqnarray}
a_1 (x) = \frac{p_{0x}}{\cosh(\alpha h)-\beta\alpha\sinh(\alpha h)},
\end{eqnarray}
\begin{eqnarray}
a_2 (x) = 0.
\end{eqnarray}
The pressure gradient $p_{0x}$ is evaluated from the constant volumetric flux condition
\begin{eqnarray}
\int_{0}^{h(x)}u_0(x,y)dy = 1,
\end{eqnarray}
and this condition yields
\begin{eqnarray} \label{e8}
p_{0x} = \frac{1}{Q(x)},
\end{eqnarray} 
where
\begin{eqnarray} \label{e12}
Q(x) = \frac{\sinh(\alpha h) - h\alpha[\cosh(\alpha h) - \beta\alpha \sinh(\alpha h)]}{\alpha[\cosh(\alpha h) - \beta\alpha\sinh(\alpha h)]}.
\end{eqnarray}
Equation \eqref{e8} shows that the total flow rate across the wavy channel is constant.
Additionally, by using the equation \eqref{mcon1} and the boundary conditions \eqref{bdcn1}-\eqref{bdcn2}, one can determine the normal velocity component as follows:
\begin{eqnarray}
 v_0(x,y)=y p_{0xx}-\left(\frac{d a_1(x)}{d x}\right)\left(\frac{\sinh(\alpha y)}{\alpha}\right)+ b_1(x).   
\end{eqnarray}
Where $b_1(x)$ is given by 
\begin{eqnarray}
 b_1(x)=-h p_{0xx}+\left(\frac{d a_1(x)}{d x}\right)\left(\frac{\sinh(\alpha h)}{\alpha}\right).   
\end{eqnarray}


\subsection{The $O(\delta^2)$ problem}
Substituting \eqref{e6} into \eqref{e2} - \eqref{e20} and equating the coefficient of $O(\delta^2)$ to zero gives the $O(\delta^2)$ problem governed by the following set of equations:
\begin{eqnarray}\label{heq1}
\alpha^2 u_1 + \alpha^2\frac{\partial p_1}{\partial x} - \frac{\partial^2 u_1}{\partial y^2} - \frac{\partial^2 u_0}{\partial x^2} = 0, 
\end{eqnarray}
\begin{eqnarray}\label{heq2}
\lambda^2\alpha^2 v_0 + \alpha^2\frac{\partial p_1}{\partial y} - \frac{\partial^2 v_0}{\partial y^2} = 0.
\end{eqnarray}

\begin{eqnarray}\label{mcon2}
\frac{\partial u_1}{\partial x} + \frac{\partial v_1}{\partial y} = 0.
\end{eqnarray}
Eliminating the pressure gradient from the first two equations \eqref{heq1} and \eqref{heq2} yields
\begin{eqnarray}
\frac{\partial}{\partial y}\left(\frac{\partial^2u_1}{\partial y^2} - \alpha^2u_1\right) = -\frac{\partial}{\partial x}(\lambda^2\alpha^2v_0) + 2\frac{\partial}{\partial y}\left(\frac{\partial^2v_0}{\partial x\partial y}\right).
\end{eqnarray}
Similarly, substituting \eqref{e6} into \eqref{e3} - \eqref{e5}	 gives the boundary conditions corresponding to the $O(\delta^2)$ problem as follows:
\begin{eqnarray} \label{e9}
u_1(x,h(x)) = \beta u_{1y}, ~\frac{\partial u_1}{\partial y} \vline_{y = 0} = 0, v_1(x,h(x))=0, v_1(x,0)=0.
\end{eqnarray} 
In addition, the corresponding volumetric flow rate condition for that problem is given by 
\begin{eqnarray}\label{e10}
\int_{0}^{h(x)}u_{1}dy = 0.
\end{eqnarray}
The general solution of $O(\delta^2)$ problem is given by
\begin{eqnarray} \label{e14}
u_1(x,y) = c_1(x)\cosh(\alpha y) + c_2(x)\sinh(\alpha y) + c_3(x)+M(x,y),
\end{eqnarray}
where $c_1(x), ~c_2(x)$ and $c_3(x)$ are functions of integration to be determined using the boundary conditions \eqref{e9} and the volumetric flux rate condition \eqref{e10}, 
\begin{eqnarray}
M(x,y) = \frac{1}{2\alpha^2}\left(f_1(x)\frac{y\sinh(\alpha y)}{2\alpha} - \frac{f_2(x)}{\alpha^2}(4-\lambda^2 \alpha^2 y^2 - 2 \lambda^2) + f_3(x)\frac{y}{\alpha^2}\right), 
\end{eqnarray}
\begin{eqnarray}
f_1(x) = 2\alpha^2(\lambda^2-2)\frac{d^2a_1(x)}{dx^2},
\end{eqnarray}
\begin{eqnarray}	
f_2(x) = \alpha^2 p_{0xxx},
\end{eqnarray}
\begin{eqnarray}
f_3(x) = 2\lambda^2\alpha^4\left(\frac{db_1(x)}{dx}\right),
\end{eqnarray}
\begin{eqnarray}
N(x,y) = \frac{\partial M}{\partial y} = \frac{1}{2\alpha^2}\left(\frac{f_1(x)\sinh(\alpha y)}{2\alpha} + \frac{y f_1(x)\cosh(\alpha y)}{2} + 2\lambda^2 yf_2(x) + \frac{f_3(x)}{\alpha^2}\right), 
\end{eqnarray}
\begin{eqnarray}
c_1(x) = \frac{\alpha^2 k(x)+N(x,0) - N(x,0)\cosh(\alpha h) + \xi_1(x) + \xi_2(x)}{\xi_3(x)},
\end{eqnarray}
\begin{eqnarray}
c_2(x) = -\frac{N(x,0)}{\alpha},
\end{eqnarray}
\begin{eqnarray}
c_3(x) = \frac{-\alpha^2 k(x)\cosh(\alpha h) + N(x,0)[1 - \cosh(\alpha h) + \alpha \beta\sinh(\alpha h)] + \xi_4}{\xi_3},
\end{eqnarray}
and
\begin{eqnarray}
\xi_1(x) = \alpha h N(x,0)[\sinh(\alpha h)-\beta \alpha\cosh(\alpha h)],
\end{eqnarray}
\begin{eqnarray}
\xi_2(x) = \alpha^2\beta h N(x,h)-\alpha^2hM(x,h),
\end{eqnarray}
\begin{eqnarray}
\xi_3 = \alpha[\alpha h \cosh(\alpha h)-\alpha^2 h \beta \sinh(\alpha h) - \sinh(\alpha h)],
\end{eqnarray}
\begin{eqnarray}
k(x) = \int_{0}^{h(x)}M(x,y)dy,
\end{eqnarray}
\begin{eqnarray}
\xi_4 = - \alpha \beta N(x,h)\sinh(\alpha h) + \alpha M(x,h)\sinh(\alpha h) + \beta k(x) \alpha^3 \sinh(\alpha h).
\end{eqnarray}
In addition, the normal velocity component $v_1(x,y)$ can be determined easily using the equation \eqref{mcon2} and the boundary conditions \eqref{e9}.

Finally, neglecting the four and higher order terms of $\delta$ in the relation \eqref{e6}, the required general solution for $u(x, y)$ of the hydrodynamic problem upto order $O(\delta^2)$  is calculated using the relations \eqref{e13} and \eqref{e14} as
\begin{eqnarray}
u(x, y) = u_0(x, y) + \delta^2 u_1(x, y).
\end{eqnarray}
The behavior of velocity profiles of the flow system will be provided by the above expression. With the above $O(\delta^0)$ and $O(\delta^2)$ flow fields, the streamlines are drawn by introducing stream functions:\\
\begin{eqnarray}
u = \psi_y, ~v = -\psi_x.
\end{eqnarray}
The stream functions are also expanded in $\delta^2$ in a similar manner like \eqref{e6}. We obtain the streamfunctions at each order by integrating the corresponding axial velocities with respect to y.

\section{Flow behaviors away from the wavy wall}\label{sec4}
Numerous studies over the past few decades have explored the flow behavior near a wall, where viscous forces play a significant role \cite{vf1, vf2, vf3}. Brinkman \cite{brinkman1949calculation} was the first to propose the notion of taking the boundary layer type equation into consideration and introduced the Brinkman equations along with the Darcy term. Thereafter a theoretical justification was presented by Tam and Lundgren \cite{tam1969drag, lundgren1972slow}. The limitation of the Brinkman equations was first introduced by Taylor \cite{taylor1971model} and thereafter by Nield and Bejan \cite{nield1983boundary}. Recently, Hill and Straughan \cite{hill2008poiseuille} examined the Poiseuille flow in a fluid covering a porous medium and introduced the existence of the Brinkman layer close to the boundary and the Darcy layer beyond. According to these researches, the flow fulfills the Brinkman equations near the wavy wall but viscous forces are insignificant away from the wall and therefore the flow obeys the Darcy equation. The corresponding Darcy equation can be obtained from the Brinkman equations by dropping the viscous term, i.e. the second-order derivative term. Hence, we have the undermentioned equations for the leading order.
\begin{eqnarray}
\alpha^2 u_0 \sim -\alpha^2 p_{0x} 
\end{eqnarray}
i.e.,
\begin{eqnarray}
u_0 \sim  -p_{0x},
\end{eqnarray}
and
\begin{eqnarray}
v_0 \sim p_{0xx} y.
\end{eqnarray}
We can determine the unknown pressure gradient using the volumetric flow rate balance as follows :
\begin{eqnarray} \label{e15}
p_{0x} \sim -\frac{1}{h(x)}.
\end{eqnarray} 
For very low permeability or large $\alpha$, we have $\alpha \gg 1$. Hence from equation
\eqref{e12} we may get
\begin{eqnarray} \label{e11}
Q(x) = \frac{\tanh(\alpha h(x))}{\alpha[1 - \alpha \beta \tanh(\alpha h(x)]} - h(x).
\end{eqnarray}
Equation \eqref{e11} may be written as
\begin{eqnarray} \label{70}
Q(x) = \frac{1}{\alpha} \times \frac{\tanh(\alpha h(x))}{[1 - \alpha \beta \tanh(\alpha h(x)]} - h(x).
\end{eqnarray}
Also since
\begin{eqnarray}
|\tanh(\alpha h(x))|\leq 1,
\end{eqnarray}
hence, for $\alpha \gg 1$, and slip coefficient $\beta$ we have
 \begin{eqnarray}
  \frac{\tanh(\alpha h(x))}{[1 - \alpha \beta \tanh(\alpha h(x)]} \le 1. 
  \end{eqnarray}
  Consequently, from equation \eqref{e8} we may get
\begin{eqnarray} \label{e16}
p_{0x} = - \frac{1}{ h(x)}.
\end{eqnarray}
Hence for the slip flow, equations \eqref{e15} and \eqref{e16} demonstrate that away from the wavy wall, more closely around the channel's center, the flow exhibits Darcy behavior. This is consistent with the porous media with low permeability.

Similarly, one can obtain the $O(\delta^2)$ velocity as
\begin{eqnarray}
u_1 \sim \lambda^2 \frac{p_{0xxx}}{2} y^2 + d_1(x).
\end{eqnarray}
The constant volumetric flow rate condition \eqref{e10} determines the value of $d_1(x)$ as 
\begin{eqnarray}
d_1(x)\sim - \lambda^2 \frac{p_{0xxx}}{6} h(x)^2.
\end{eqnarray}
The axial pressure gradient can be obtained from 
\begin{eqnarray}
\frac{\partial p}{\partial x} \sim \frac{\partial p_0}{\partial x} + \delta^2 \frac{\partial p_1}{\partial x}.
\end{eqnarray}
As a result, we can obtain the axial pressure gradient corresponding to Darcy's equation as 
\begin{eqnarray}
\frac{\partial p}{\partial x} \sim \frac{\partial p_0}{\partial x} - \delta^2\lambda^2\frac{p_{0xxx}}{2}\left(y^2 - \frac{h(x)^2}{3}\right).
\end{eqnarray}
Similarly, one can determine the corresponding axial velocity as
\begin{eqnarray}
u \sim - \frac{\partial p_0}{\partial x} + \delta^2\lambda^2 \frac{p_{0xxx}}{2}\left(y^2 - \frac{h(x)^2}{3}\right).
\end{eqnarray}

\section{Discussion on analytical results}\label{sec5}

The governing model with the presence of slip boundary conditions is solved analytically via the perturbation technique \cite{ng2010darcy,  tsangaris1984laminar,  karmakar2017note,  wei2003flow,luo2008two}, and finally,  in this section, the obtained expressions for velocity and wall pressure are used for graphical representation of the flow dynamics for the considered model. 
The analytical results are graphically presented for a wide range of parameter values, obtained using a MATLAB code developed to examine the influence of wall velocity slip $(\beta)$, anisotropic ratio $(\lambda)$, porosity parameter $(Da)$, and other relevant factors on various hydrodynamic quantities, including axial velocity, shear stress, and streamlines.

\begin{figure} [t]
\centering
\includegraphics[height= 6 cm]{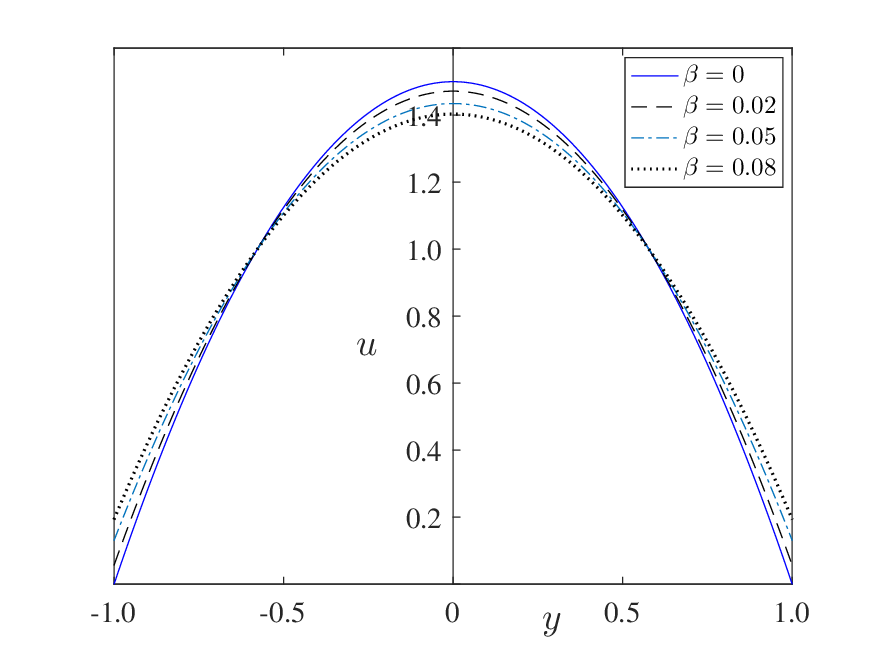}
\caption{Effect of the wall slip parameter $\beta$ on the axial velocity profile of the corresponding plane Poiseuille flow at $x$ = 0 for $\delta$ = 0, $a$ = 0, $\lambda = 1$ $Da=10^{35}$.}
\label{Threshold_fig 122}
\end{figure}	

\begin{figure}[b]
 \centering
 \subfigure[]{\label{fig:3a}\includegraphics*[width=7cm]{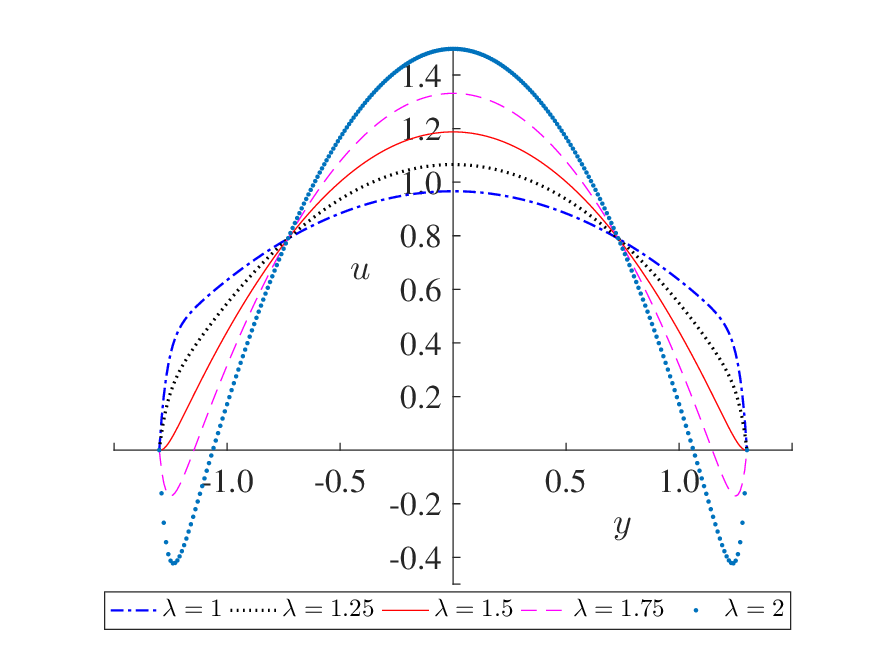}}\subfigure[]{\label{fig:3b}\includegraphics*[width=7cm]{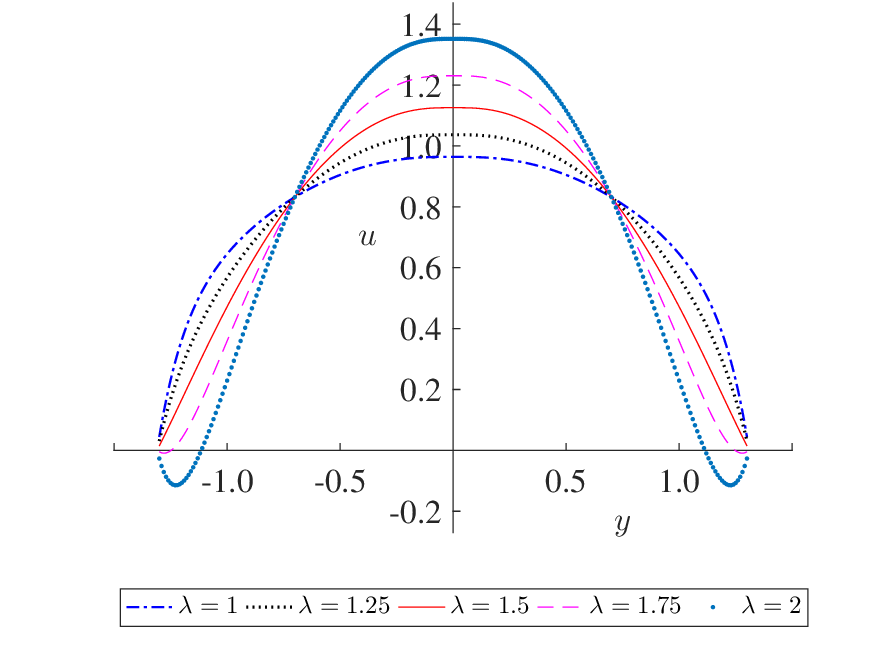}}
 \end{figure}
 \begin{figure}[t]
 \centering
 \subfigure[]{\label{fig:3c}\includegraphics*[width=7cm]{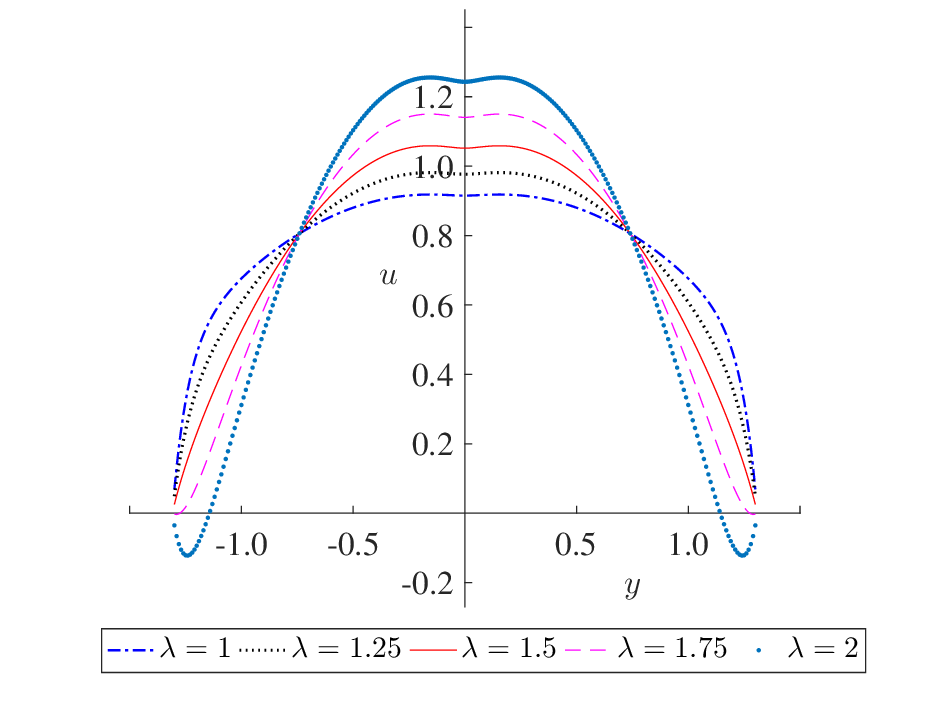}}\subfigure[]{\label{fig:3d}\includegraphics*[width=7cm]{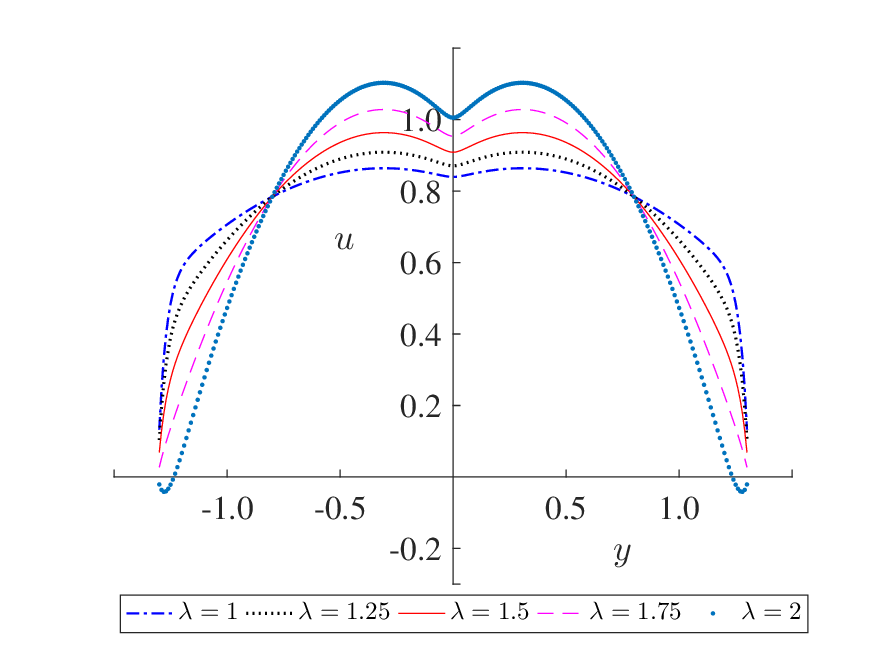}}
 \caption{The impact of anisotropic ratio $\lambda$ on the axial velocity profile $u(x,y)$ at $x$ = 0 for $\delta$ = 0.3, $a$ = 0.3, and (a) without velocity slip ($\beta$ = 0.0), \& $Da = 10^{-3}$, (b) with velocity slip ($\beta$ = 0.01) \& $Da = 10^{-2}$, (c) with velocity slip ($\beta$ = 0.01) \& $Da = 4 \times 10^{-3}$, (d)  with velocity slip ($\beta$ = 0.01) \& $Da = 10^{-3}$.} \label{f3}
 \end{figure}

\subsection{Axial velocity profile at the crest of the wavy wall:}
Figure \ref{Threshold_fig 122} displays the effect of the slip parameter $\beta$ on the axial velocity profile $u(x,y)$ for $a$ = 0, $\delta$ = 0, and a higher value of the porosity parameter $Da$. Notably, this figure represents the dynamics corresponding to a plane Poiseuille flow. The behavior of the axial velocity profile has been discussed by taking the various values of the dimensionless slip parameters as $\beta=0, 0.02, 0.05, 0.08$. Figure \ref{Threshold_fig 122} illustrates that the increment of the slip parameter $\beta$ increases wall velocity while decreasing velocity near the centerline and the velocity profile becomes progressively flatter, approaching plug flow behavior. 
It is observed that all findings related to plane Poiseuille flow are consistent with those reported in \textcolor{red}{[Put suitable reference]}.
Building on the agreement with existing results, the analysis is extended to investigate flow behavior in the porous medium, considering the effects of wall slip on wavy walls. Figures \ref{fig:3a}, \ref{fig:3b}, \ref{fig:3c} and \ref{fig:3d} demonstrate the scaled axial velocity profiles for three distinct values of the Darcy number, $Da=10^{-3}, 4 \times 10^{-3},$ and $ 10^{-2}$, respectively. In each case, we study the effect of the anisotropic ratio $\lambda$ on the scaled axial velocity profile for two configurations: (i) without velocity slip in Figure \ref{fig:3a} ($\beta = 0$) and (ii) with velocity slip ($\beta = 0.01$) at the wavy walls in Figures \ref{fig:3b}, \ref{fig:3c} and \ref{fig:3d}. The findings presented in Figure \ref{fig:3a} represent a limiting case of the previous study by Karmakar et al. \cite{karmakar2017note} and yield identical results. These figures demonstrate that increasing the anisotropic ratio $\lambda$ leads to faster flow near the centerline and slower velocity near the walls.
Physically, an increase in the anisotropic ratio $\lambda$ enhances permeability along the $x$-axis, which in turn reduces frictional forces near the centerline, leading to a potential increase in fluid velocity. In addition, from Figures  \ref{fig:3b},  \ref{fig:3c},  \ref{fig:3d}, it is observed that the presence of velocity slip at the wavy walls enhances the velocity near the walls and reduces the velocity around the centerline in order to satisfy the constant volumetric flow rate condition. In other words, the presence of wall slip initially accelerates the axial flow near the walls while significantly slowing it down toward the channel centerline. Notably, the presence of wall velocity slip and a decrease in the porosity parameter $Da$ gradually mitigate the backflow phenomenon.
Physically, this implies that wall slip plays a significant role in enhancing the axial velocity profile. Additionally, Figures \ref{fig:3a} and \ref{fig:3b} illustrate that in both cases the fluid velocity is an increasing function of $y$ corresponding to $\lambda = 1.5, 1.75, 2$ across the flow domain $ D_1=(-1.3 \le y \le 0)$ and a decreasing function across the flow domain $ D_2 = (0 \le y \le 1.3)$.
However, in deviation from the usual pattern, one part of the fluid moves in one direction while the other part flows in the opposite direction. More specifically, there is a critical value $y_c$ (approximately $-0.75$ for domain $D1$ and approximately $0.75$ for domain $ D2$) of the flow variable $y$ that separates the flow domain into two zones. In the first zone of $D1$, adjacent to the wall, the fluid velocity increases with an increase in anisotropic ratio $\lambda$, while an opposite scenario is observed in the other zone adjacent to the centerline. A similar trend is noticed in the other half of the flow domain $D_2$ due to the symmetric behavior of the flow. Moreover, Figures \ref{fig:3b}, \ref{fig:3c} and \ref{fig:3d} show that in the presence of wall slip, reducing Darcy's number alters the curvature of the velocity profile.
In addition, abrupt changes take place in the axial velocity profile towards the centerline of the flow domain. Further, in the presence of wall slip, a small Darcy number introduces strong porous resistance in the bulk while allowing finite velocity at the wall. This mismatch causes a rapid adjustment of velocity from the wall to the center, resulting in an abrupt change in the velocity profile. Consequently, the physical impacts of the porosity parameters $Da$ on the velocity profile are visualized.
\begin{figure}[t]
 \centering
 \subfigure[]{\label{fig:avse}\includegraphics*[width=7cm]{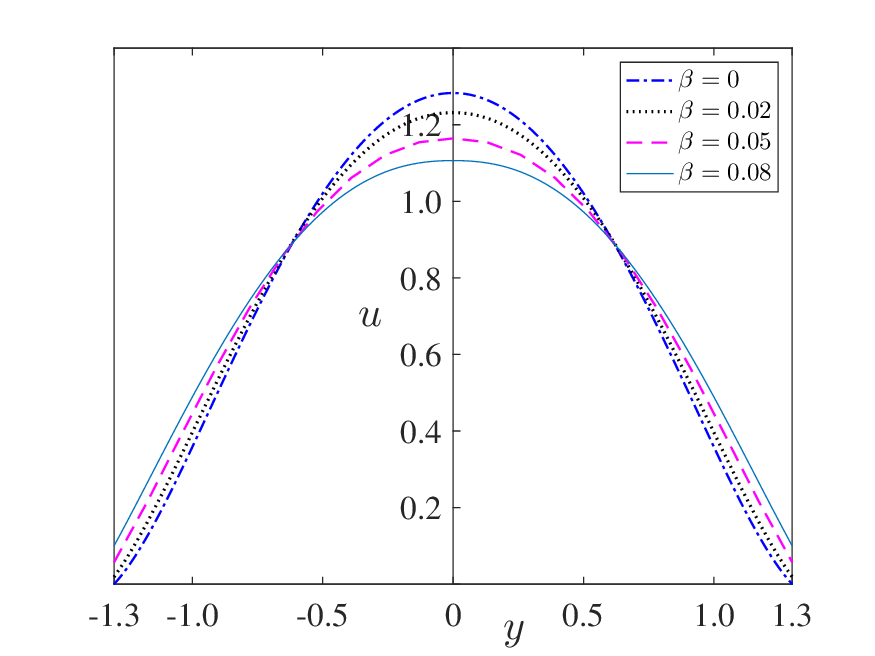}}\subfigure[]{\label{avse1}\includegraphics*[width=7cm]{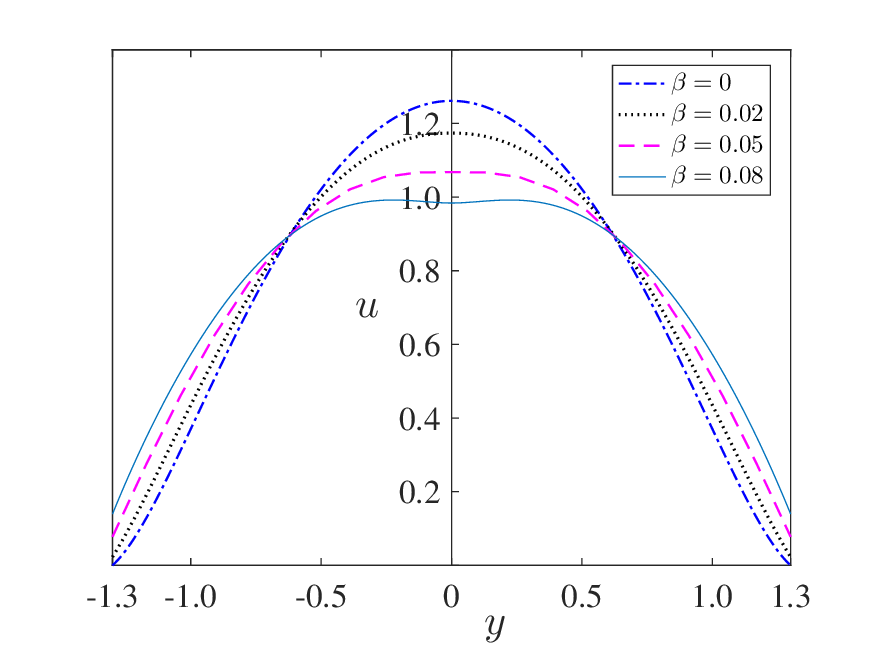}}
 \caption{Effect of slip parameter $\beta$ on the axial velocity profile at $x$ = 0 for $\delta$ = 0.3, $a$ = 0.3, $\lambda = 1.5$  and (a) $Da = 10^{-1}$, (b) $Da = 4\times 10^{-2}$.} \label{f3}
 \end{figure}

Figures \ref{fig:avse} and \ref{avse1} illustrate the effect of slip parameters $\beta$ on the axial velocity for two distinct Darcy numbers (Figure \ref{fig:avse} with $Da = 10^{-1}$, Figure \ref{avse1} with $Da = 4 \times 10^{-2}$), while the other flow parameter values are fixed. In each figure, the slip parameter varies as $\beta = 0, 0.2, 0.5, 0.8$.
It is observed from both figures that an increase in the slip parameter $\beta$ allows more fluid particles to move toward the region near the wall with reduced resistance. Consequently, this leads to an increase in wall velocity, while the velocity near the centerline decreases as the slip parameter increases. Physically, it means that stronger wall slips can overcome the viscous effects at wall zones.
 Furthermore, for a relatively small Darcy number (in Figure \ref{avse1} with $Da=4 \times 10^{-2}$), the curvature of the velocity profile changes with stronger wall velocity slip $\beta$ which alters the overall shear rate.  It is also identified that the axial velocity profile changes its behavior near the position $y = \pm 0.75$ in both cases. It is worth pointing out from these two figures that the velocity distribution corresponding to a small Darcy number becomes flatter around the centerline.

\begin{figure}[b]
		\centering
		\includegraphics[height=3.5cm]{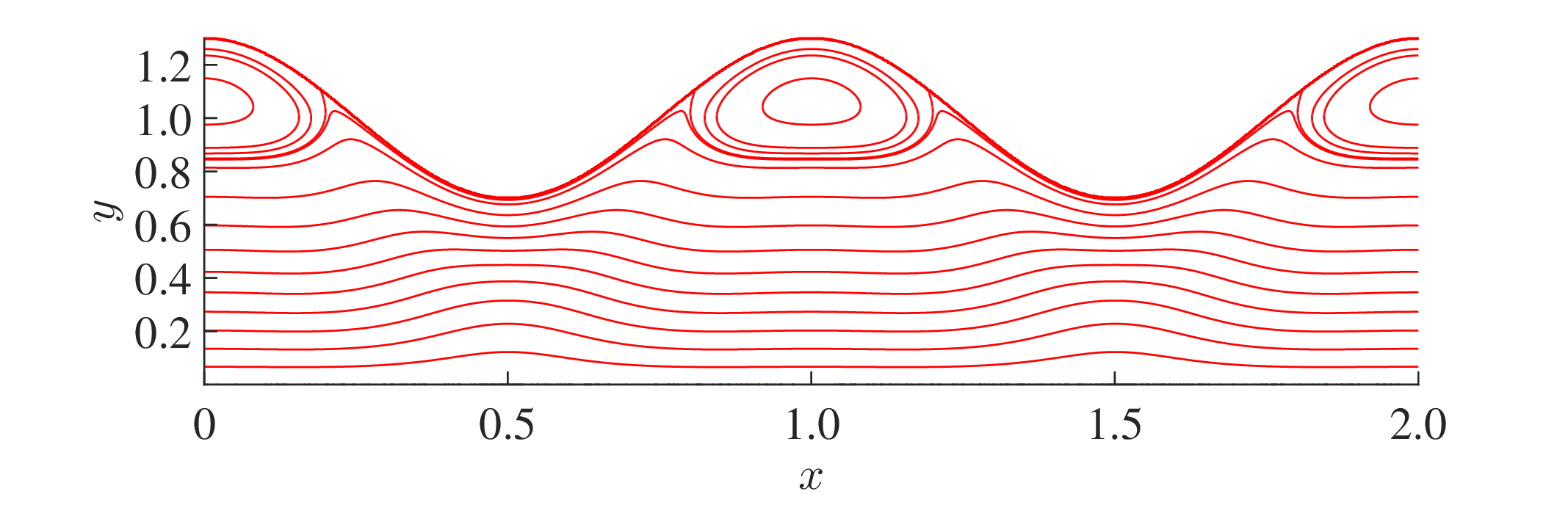}
		\caption{Streamline for $\delta$ = 0.3, $a$ = 0.3, $Da = 10^{-3}$, $\lambda$ = 2 and $\beta$ = 0.}
		\label{Threshold_fig 5}
	\end{figure}
	
	\begin{figure}[t]
		\centering
		\includegraphics[height=3.5cm]{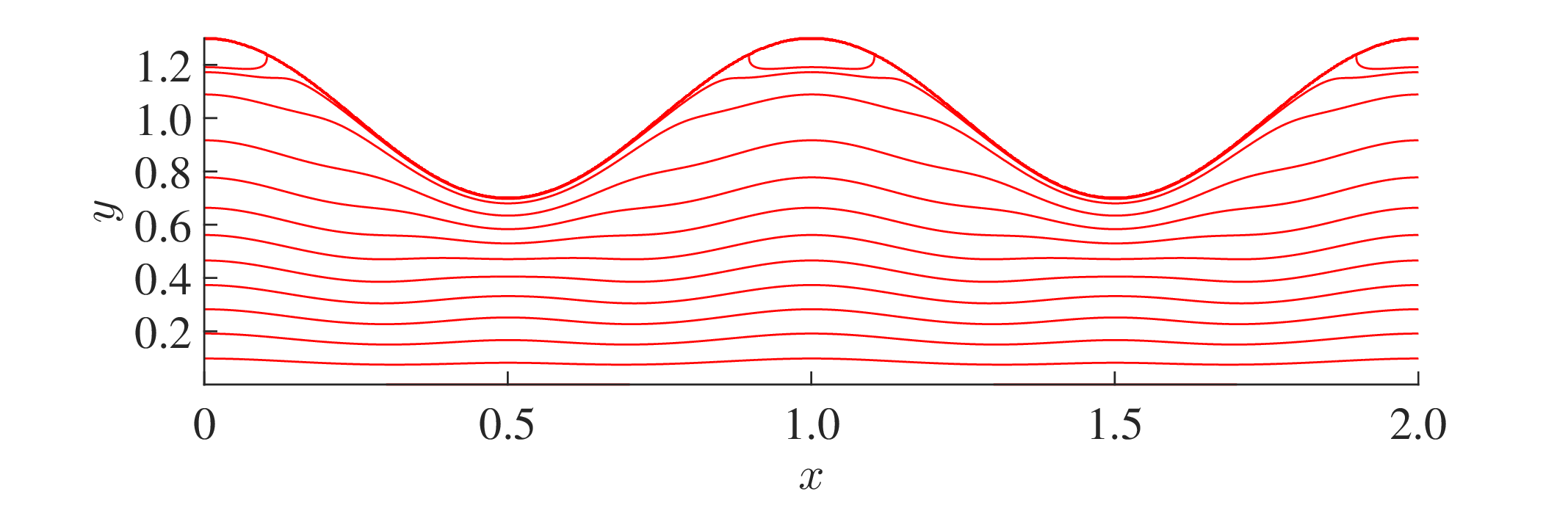}
		\caption{Streamline for $\delta$ = 0.3, $a$ = 0.3, $Da = 10^{-3}$, $\lambda$ = 2 and $\beta = 0.01$.}
		\label{Threshold_fig 6}
	\end{figure}
	
	\begin{figure}[h]
		\centering
		\includegraphics[height=3.5cm]{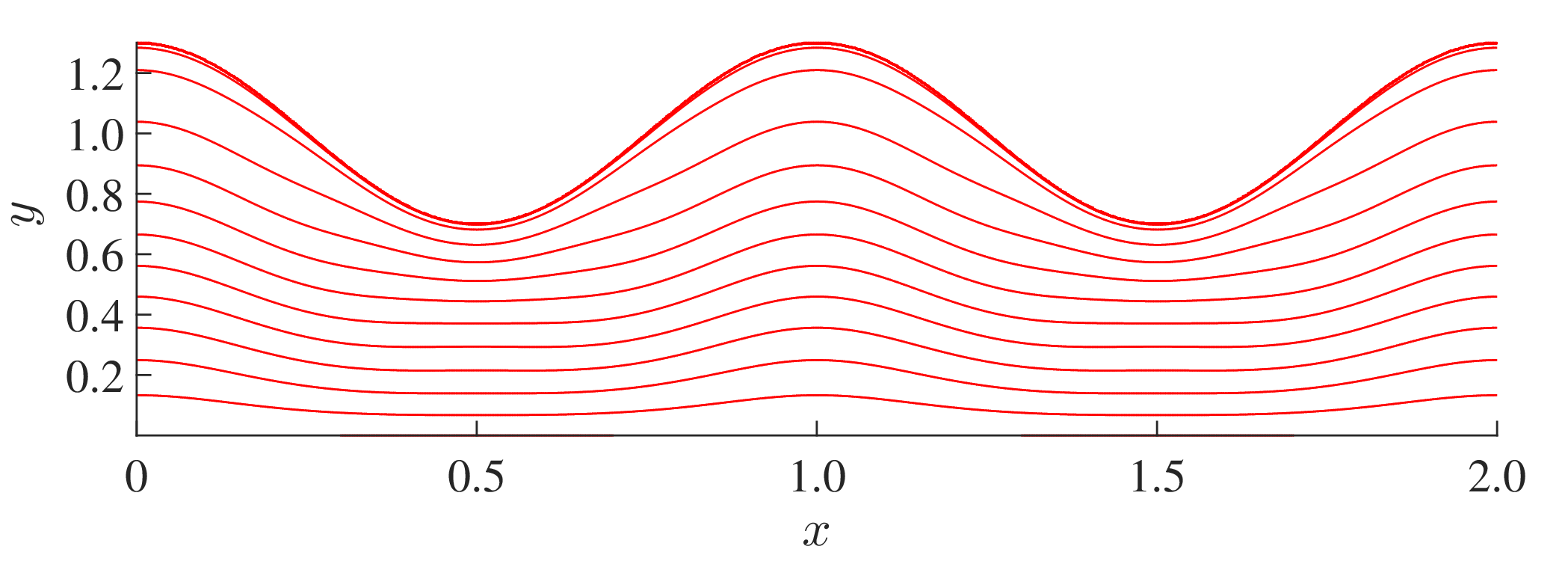}
		\caption{Streamline for $\delta$ = 0.3, $a$ = 0.3, $Da = 10^{-3}$, $\lambda$=2 and $\beta$ = 0.02.}
		\label{Threshold_fig 7}
	\end{figure}
    \subsection{Streamline variation:}
\begin{figure}[h]
		\centering
		\includegraphics[height=3.5cm]{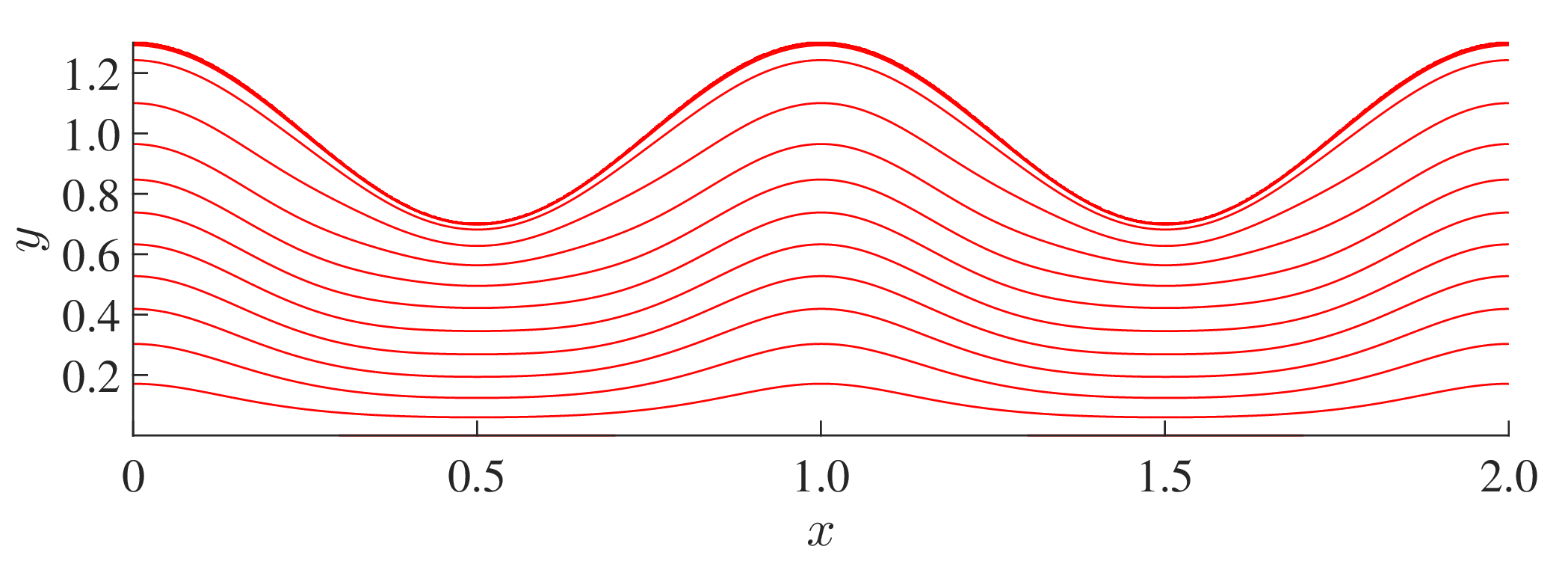}
		\caption{Streamline for $\delta$ = 0.3, $a$ = 0.3, $Da = 10^{-3}$, $\lambda$ = 2 and $\beta$ = 0.03.}
		\label{Threshold_fig 8}
	\end{figure}

\begin{figure}[h]
	\centering
	\includegraphics[height=3.5cm]{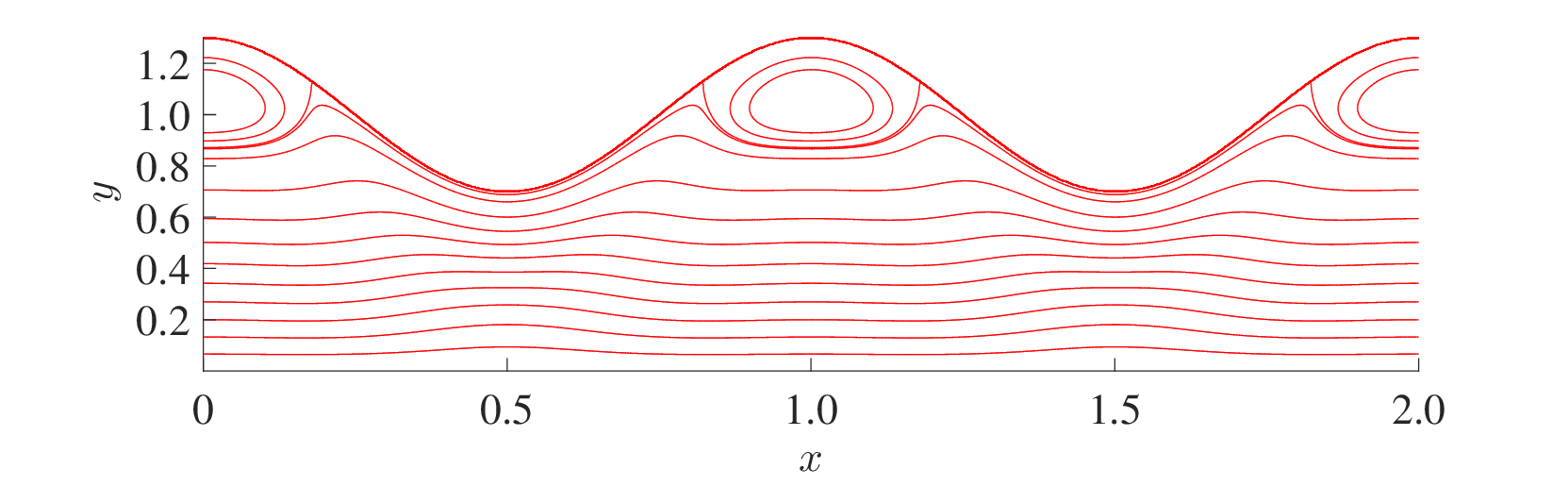}
	\caption{{Streamline for $\delta$ = 0.3, $a$ = 0.3, $Da = 10^{-2}$, $\lambda$ = 2 and $\beta$ = 0.}}
	\label{Threshold_fig 9}
\end{figure}

\begin{figure}[h]
	\centering
	\includegraphics[height=3.5cm]{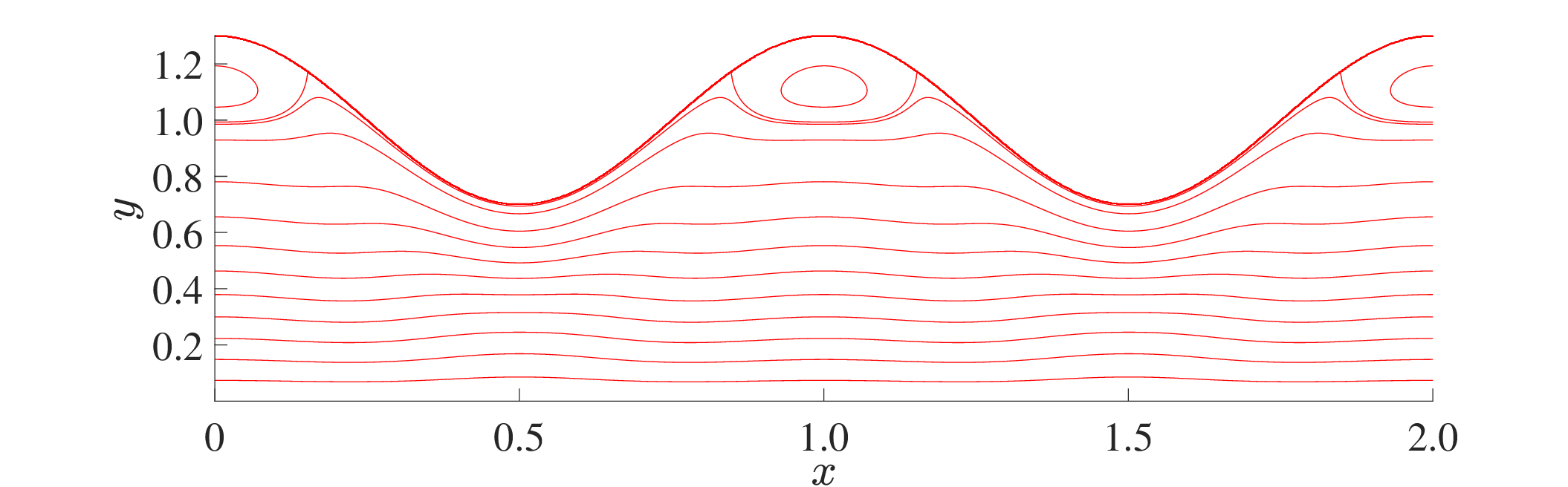}
	\caption{{Streamline for $\delta$ = 0.3, $a$ = 0.3, $Da = 10^{-2}$, $\lambda$ = 2 and $\beta$ = 0.01.}}
	\label{Threshold_fig 10}
   \end{figure}

\begin{figure}[h]
		\centering
		\includegraphics[height=3.5cm]{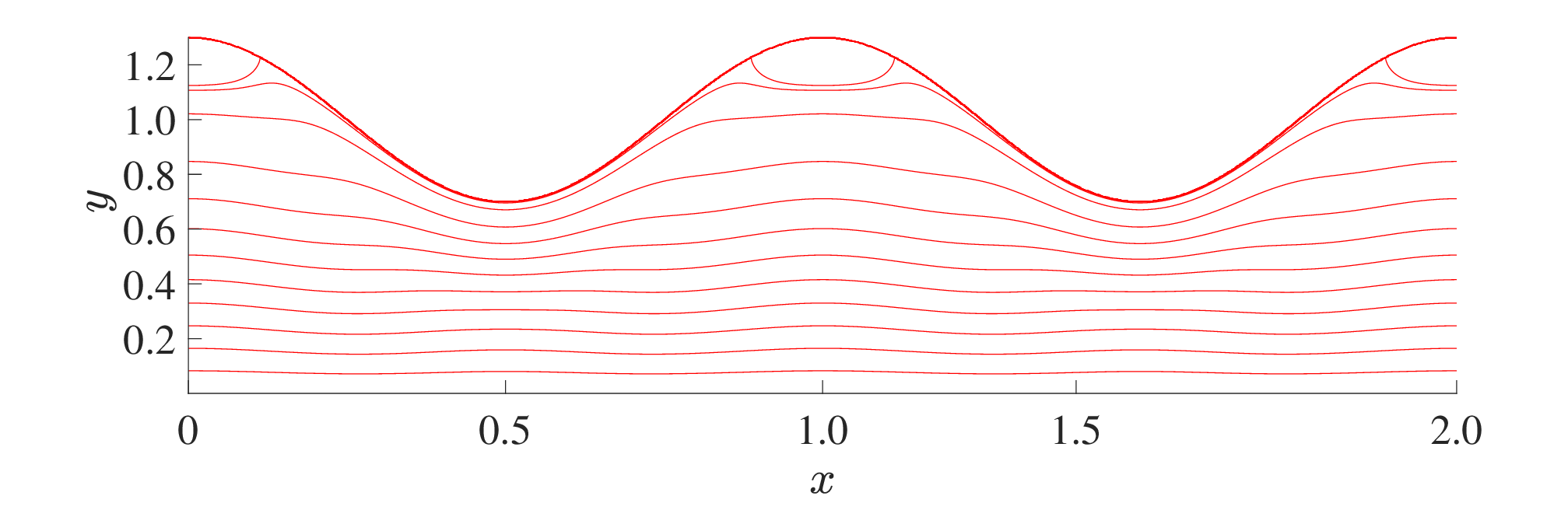}
		\caption{{Streamline for $\delta$ = 0.3, $a$ = 0.3, $Da = 10^{-2}$, $\lambda$ = 2 and $\beta$ = 0.02.}}
		\label{Threshold_fig 11}
	\end{figure}

\begin{figure}[h]
	\centering
	\includegraphics[height=3.5cm]{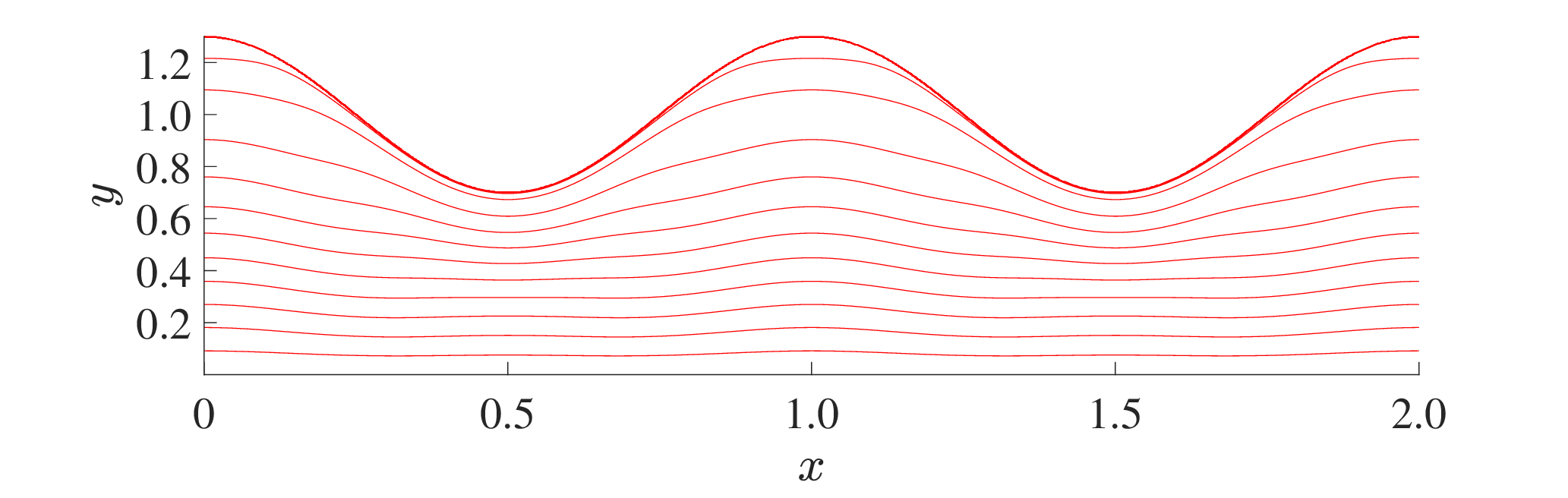}
	\caption{{Streamline for $\delta$ = 0.3, $a$ = 0.3, $Da = 10^{-2}$, $\lambda$ = 2 and $\beta$ = 0.03.}}
	\label{Threshold_fig 12}
	
\end{figure}

\begin{figure}[h]
		\centering
		\includegraphics[height=3.5cm]{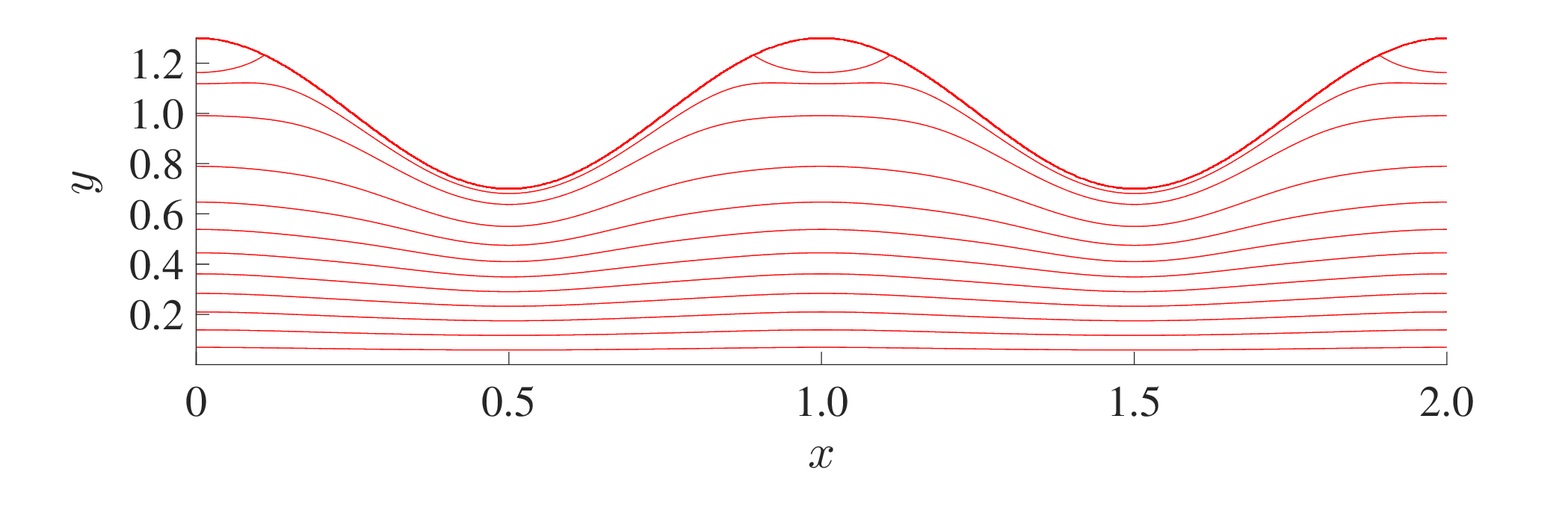}
		\caption{{Streamline for $\delta$ = 0.3, $a$ = 0.3, $Da = 10^{-1}$, $\lambda$ = 2 and $\beta$ = 0.}}
		\label{Threshold_fig 13}
     	\end{figure}

\begin{figure}[h]
	\centering
	 \includegraphics[height=3.5cm]{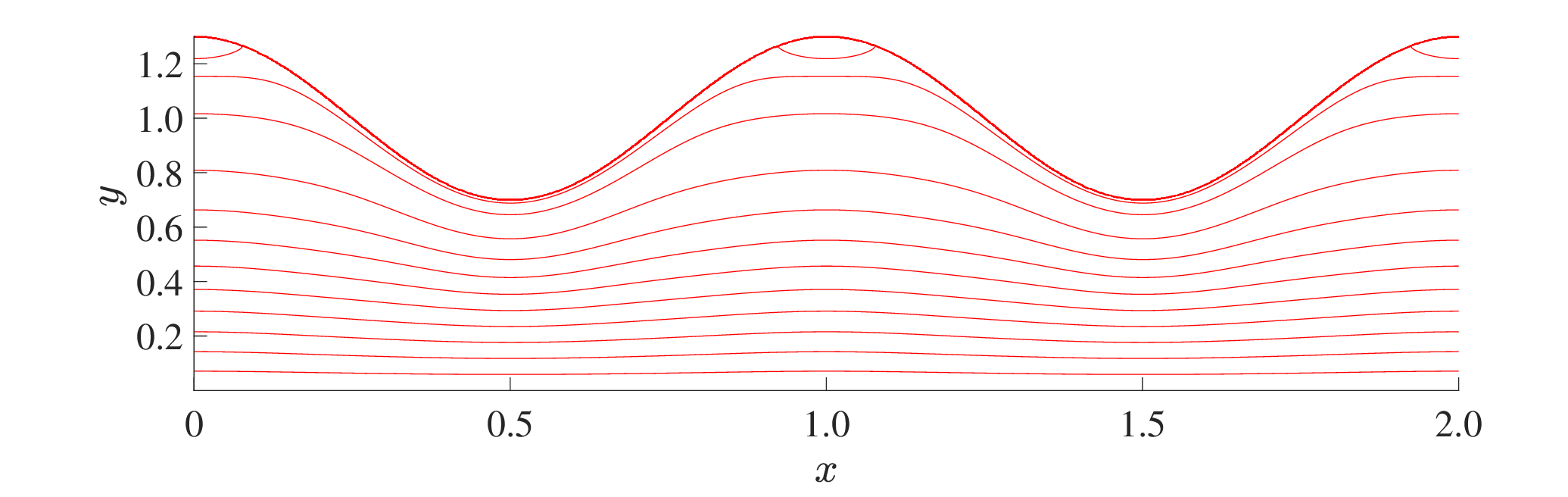}
	 \caption{{Streamline for $\delta$ = 0.3, $a$ = 0.3, $Da = 10^{-1}$, $\lambda$ = 2 and $\beta$ = 0.01.}}
	 	\label{Threshold_fig 14}
	 \end{figure}

\begin{figure}[h]
	 	\centering
	 	\includegraphics[height=3.5cm]{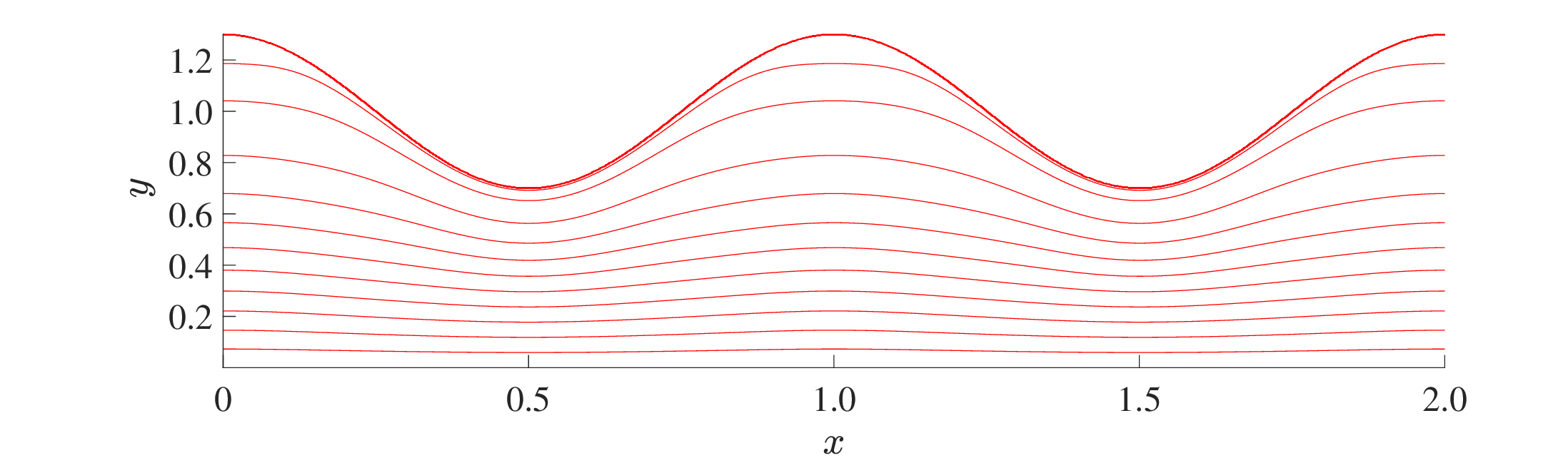}
	 	\caption{{Streamline for $\delta$ = 0.3, $a$ = 0.3, $Da = 10^{-1}$, $\lambda$ = 2 and $\beta$ = 0.02.}}
	 	\label{Threshold_fig 15}
	 \end{figure}

\begin{figure}[h]
	 	\centering
	 	\includegraphics[height=4.8cm]{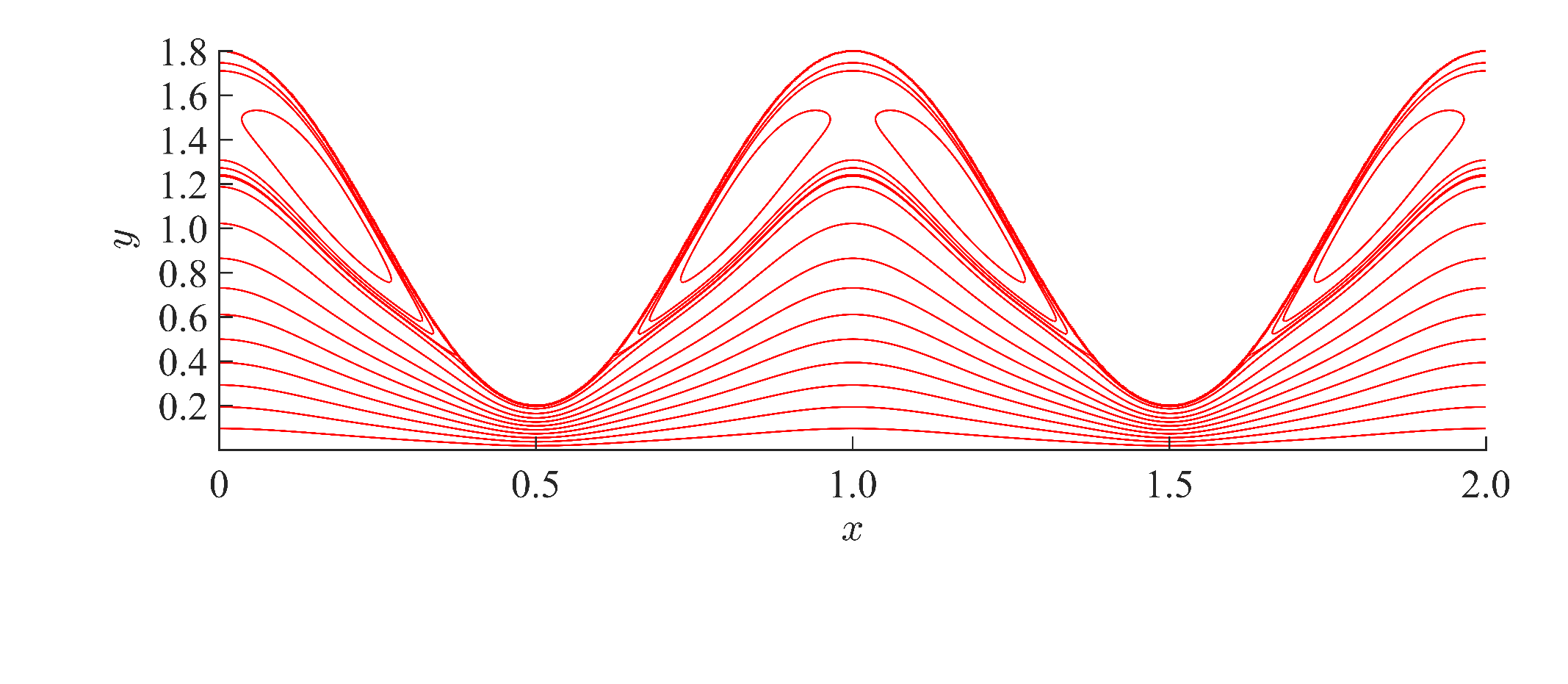}
	 	\caption{{Streamline for $\delta$ = 0.3, $a$ = 0.8, $Da = 10^{-3}$, $\lambda$ = 1 and $\beta$ = 0.0.}}
	 	\label{Threshold_fig 16}
	 \end{figure}

\begin{figure}[h]
	 	\centering
	 	\includegraphics[height=4.8cm]{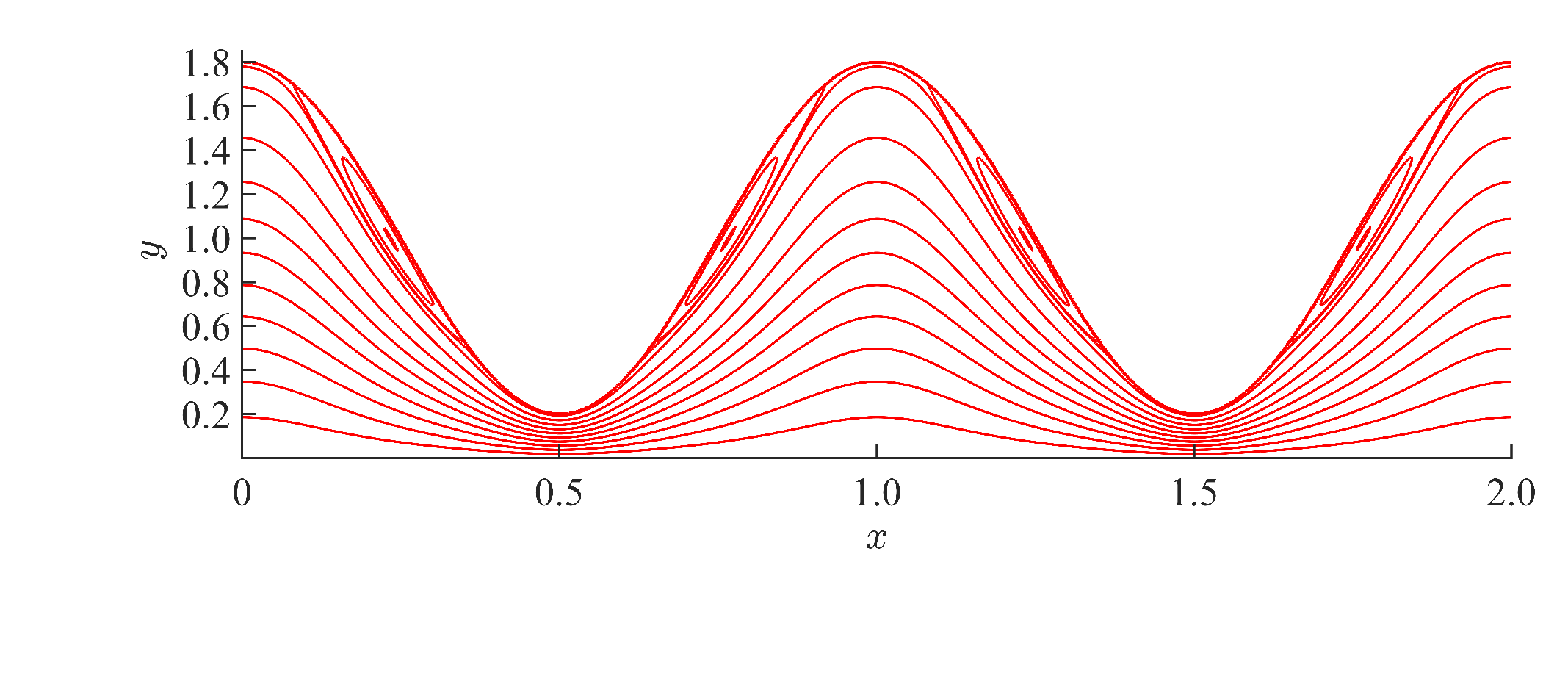}
	 	\caption{{Streamline for $\delta$ = 0.3, $a$ = 0.8, $Da = 10^{-3}$, $\lambda$ = 1 and $\beta$ = 0.02.}}
	 	\label{Threshold_fig 17}
	 \end{figure}

\begin{figure}[h]
	 	\centering
	 	\includegraphics[height=4.8cm]{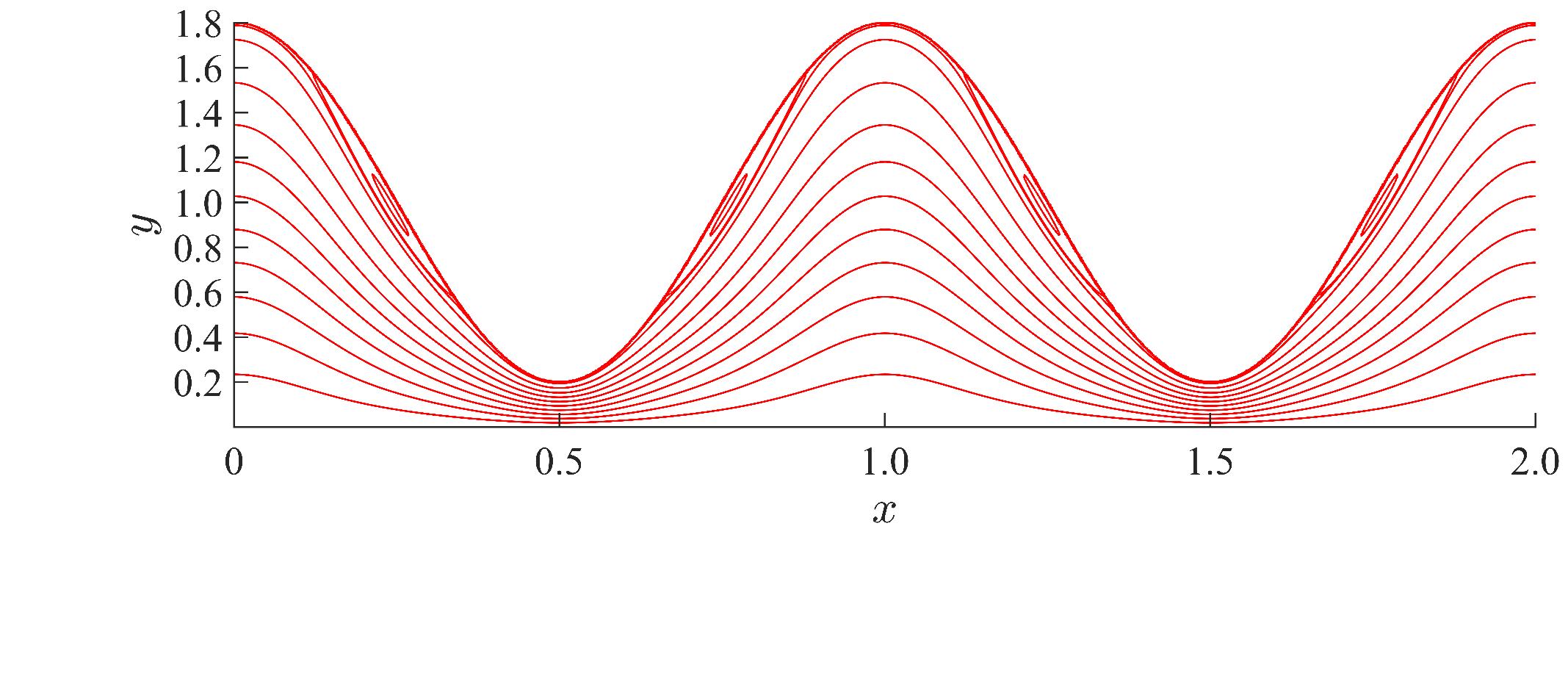}
	 	\caption{{Streamline for $\delta$ = 0.3, $a$ = 0.8, $Da = 10^{-3}$, $\lambda$ = 1 and $\beta$ = 0.03.}}
	 	\label{Threshold_fig 18}
	 \end{figure}

 The considered flow dynamics in a wavy channel are analyzed by plotting the streamline contours in Figures \ref{Threshold_fig 5}–\ref{Threshold_fig 15} for three different Darcy numbers ($Da = 10^{-3}, 10^{-2}, 10^{-1}$). In each figure, the other flow parameters are kept fixed while varying the slip parameter $\beta$. Notably, the streamlines emerge and vanish at multiple discrete points along the wall for both cases; (i) no-slip ($\beta=0$) and slip ($\beta \neq 0$) conditions. Figures \ref{Threshold_fig 5}–\ref{Threshold_fig 8} illustrate the streamline patterns for $Da=10^{-3}$. Interestingly, Figure \ref{Threshold_fig 5} with $\beta=0$ represents the limiting case of the previous study by Karmakar et al. \cite{karmakar2017note}, reproducing their results exactly. The phenomenon of flow reversal can be interpreted through axial velocity variations. A closer examination of Figure \ref{fig:3a} reveals that the velocity profile transitions from positive to negative for $\lambda=1.75$ and $\lambda=2$, signifying the occurrence of flow separation.
Figure \ref{Threshold_fig 5} demonstrates the formation of vortices near the crest of the wavy wall, where the flow starts to circulate, leading to the development of flow separation zones in the vicinity of the crest. Figures \ref{Threshold_fig 6}, \ref{Threshold_fig 7}, and \ref{Threshold_fig 8} depict several fascinating flow reversal characteristics influenced by slip effects.
For slip flow at different Darcy numbers $Da$, Figures \ref{fig:3b}, \ref{fig:3c}, and \ref{fig:3d} reveal that the velocity profile transitions from positive to negative only for $\lambda=2$, indicating the occurrence of flow reversal. Next, our aim is to examine the influence of slip parameters on flow reversal. It is observed that an increase in the slip parameter significantly diminishes the flow reversal zones. An increase in the velocity slip parameter reduces wall shear stress and preserves near-wall fluid momentum. This weakens the influence of adverse pressure gradients and prevents flow reversal. Consequently, flow variations decrease, leading to a smoother flow pattern near the crest of the wavy wall. Notably, greater slip at the wall displaces the streamline at $x=0,1,2$ farther from the channel centerline. It is noticed that the streamlines upstream initially follow the contour of the wavy wall, and with an increase in $\beta$, they gradually adapt more closely to its geometry.
Figures \ref{Threshold_fig 9}–\ref{Threshold_fig 12} depict the streamline patterns for $Da=10^{-2}$ with increasing slip parameters $\beta$, while Figures \ref{Threshold_fig 13}–\ref{Threshold_fig 15} illustrate the streamlines for $Da=10^{-1}$ under varying slip conditions. Figures \ref{Threshold_fig 5}, \ref{Threshold_fig 9}, and \ref{Threshold_fig 13} illustrate that as $Da$ increases, the flow reversal zones at the top of the wavy wall diminish in the absence of slip flow. This indicates that for a weaker porous medium $(Da=10^{-2}, 10^{-1})$ with $\beta=0$, the flow remains relatively smooth. Interestingly, increasing the slip parameter in a stronger porous medium ($Da=10^{-3}$) suppresses flow reversal zones more rapidly compared to a weaker porous medium ($Da=10^{-2}, 10^{-1}$).Thus, significant findings emerge when slip parameters are introduced in both stronger and weaker porous media. Physically, these results indicate that slip parameters have a dynamic influence on fluid motion, particularly affecting the trapping phenomenon in varying porous media strengths. Furthermore, Figures
\ref{Threshold_fig 16}, \ref{Threshold_fig 17} and  \ref{Threshold_fig 18} show the effect of different slip parameter values in isotropic porous media with $\lambda=1$ on the streamline variations, while the other parameters are fixed at $\delta = 0.3$, $Da = 10^{-3}$, and $a = 0.08$. Additionally, Figure \ref{Threshold_fig 16} illustrates the flow recirculation at the crests $x=0,1$, and $2$ in the absence of slip. Notably, flow variability is more pronounced compared to cases with a smaller amplitude parameter. An increase in the amplitude parameter intensifies flow disturbance and nonlinear interactions, leading to enhanced spatial and temporal variations in the velocity field and, consequently, greater flow variability. In an isotropic porous medium, an increase in the amplitude parameter intensifies flow acceleration and deceleration, leading to strong adverse pressure gradients in the expansion regions. When these gradients overcome the forward momentum of the fluid, flow reversal occurs. Further, It is also observed that the size of the vortices varies which are created depending on the potential of the backflow configurations. Figures \ref{Threshold_fig 17} and \ref{Threshold_fig 18} display the streamlines with slip values $\beta = 0.02$ and $0.03$. From these figures, it is seen that the flow reversal decreases as the slip parameter increases, which results in small vortices. For high amplitude parameters, although strong adverse pressore gradients tend to promote flow reversal, an increase in the velocity slip parameter reduces wall shear stress and preserves near-wall momentum. In an isotropic porous medium, the uniform resistance further stabilizes the flow, leading to a reduction in flow reversal. Additionally, wall slip causes the streamline to move away from the centerline at $x = 0, 1, 2$.



 \begin{figure}[t]
 \centering
 \subfigure[]{\label{fig:8a}\includegraphics*[width=7cm]{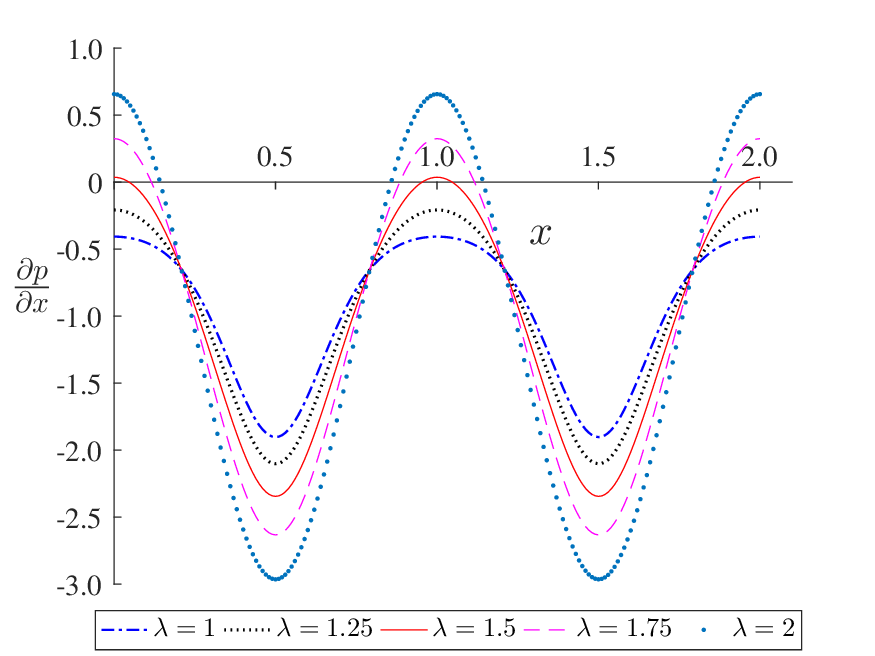}}\subfigure[]{\label{fig:8c}\includegraphics*[width=7cm]{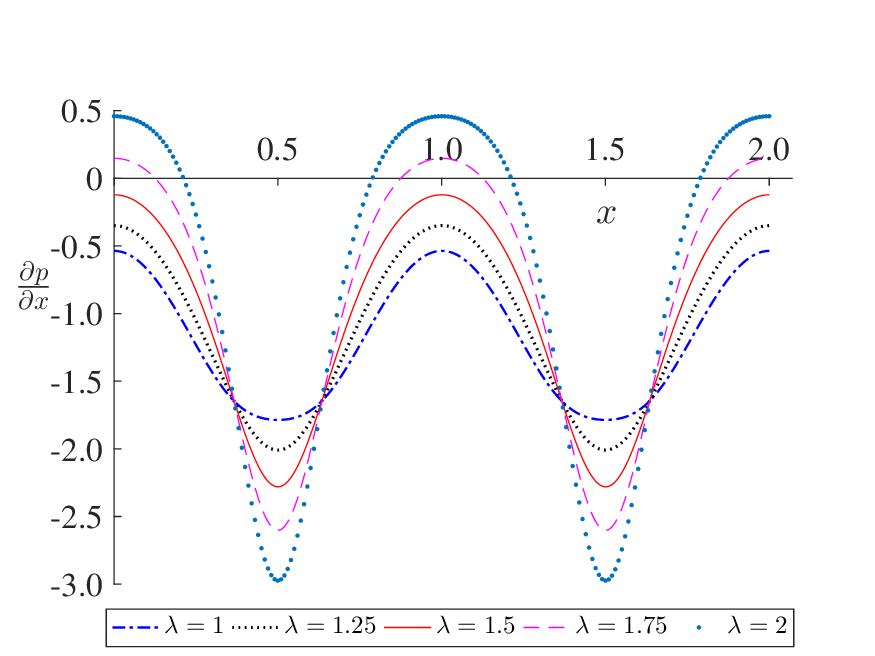}}
 \end{figure}

 \begin{figure}[ht!]
 \centering
 \subfigure[]{\label{fig:8d}\includegraphics*[width=7cm]{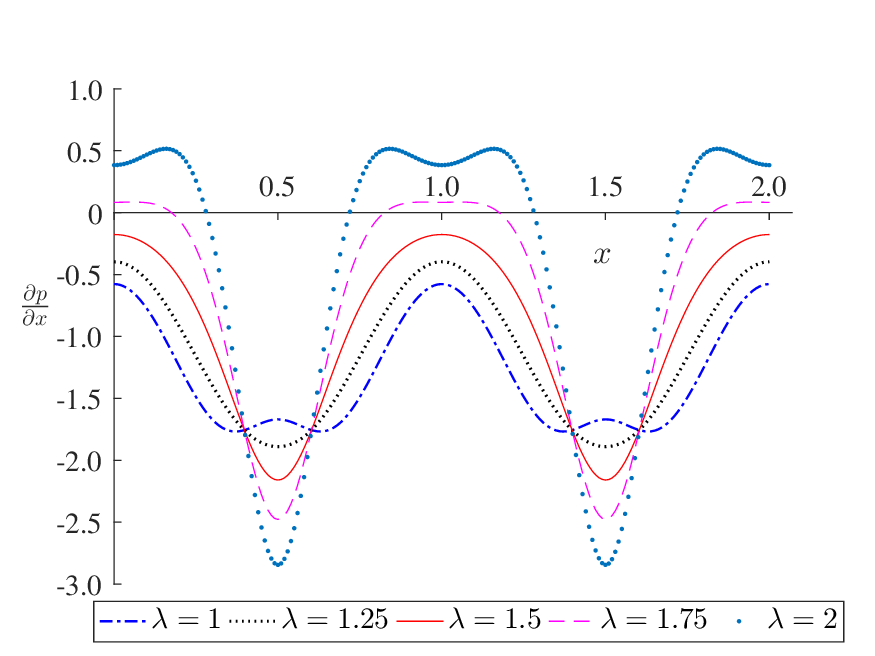}}\subfigure[]{\label{fig:8b}\includegraphics*[width=7cm]{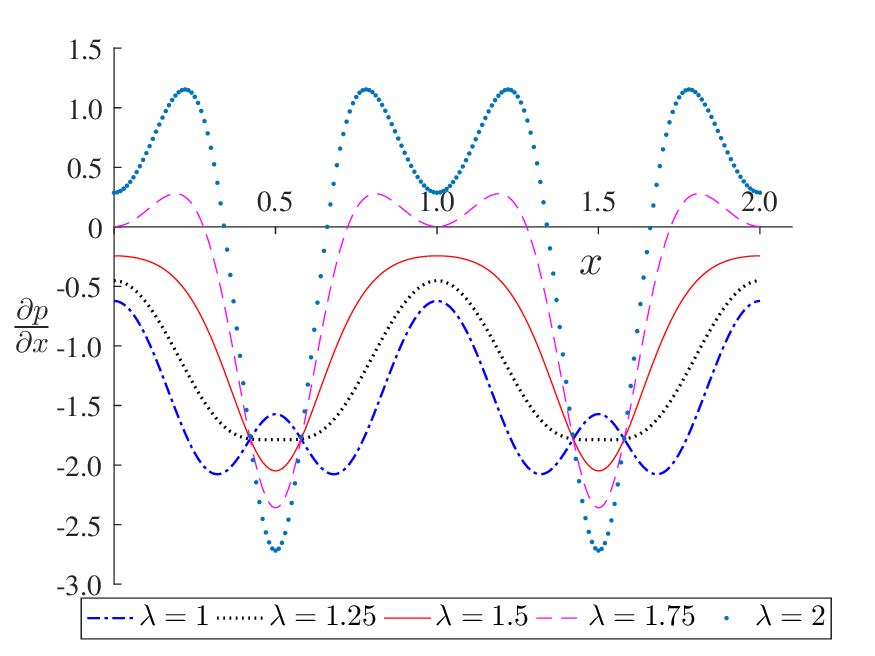}}
 \caption{Impact of $\lambda$ on the wall pressure gradient for $\delta$ = 0.3, $a$ = 0.3, (a) without velocity slip ($\beta$ = 0.0) and  $Da = 10^{-3}$,  (b) with velocity slip ($\beta$ = 0.01) and $Da = 4 \times10^{-3}$, (c) with velocity slip ($\beta$ = 0.01) and $Da = 2 \times 10^{-3}$, (d) with velocity slip ($\beta$ = 0.01) and $Da = 10^{-3}$.} \label{f3}
 \end{figure}
 \subsection{Wall pressure distribution:}
\begin{figure}[t]
 \centering
 \subfigure[]{\label{fig:wpslip}\includegraphics*[width=7cm]{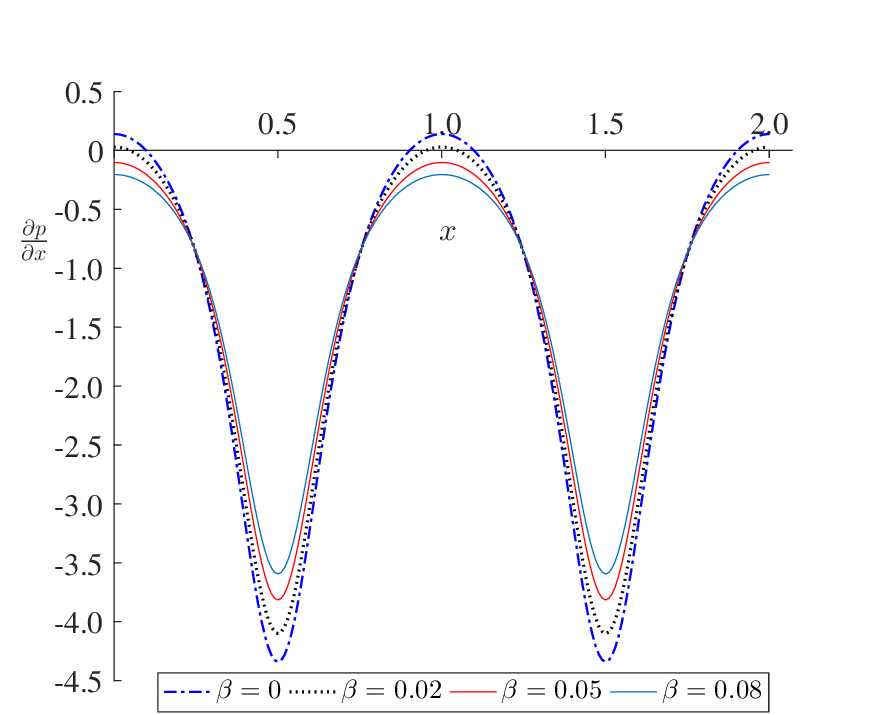}}\subfigure[]{\label{wpslip1}\includegraphics*[width=7cm]{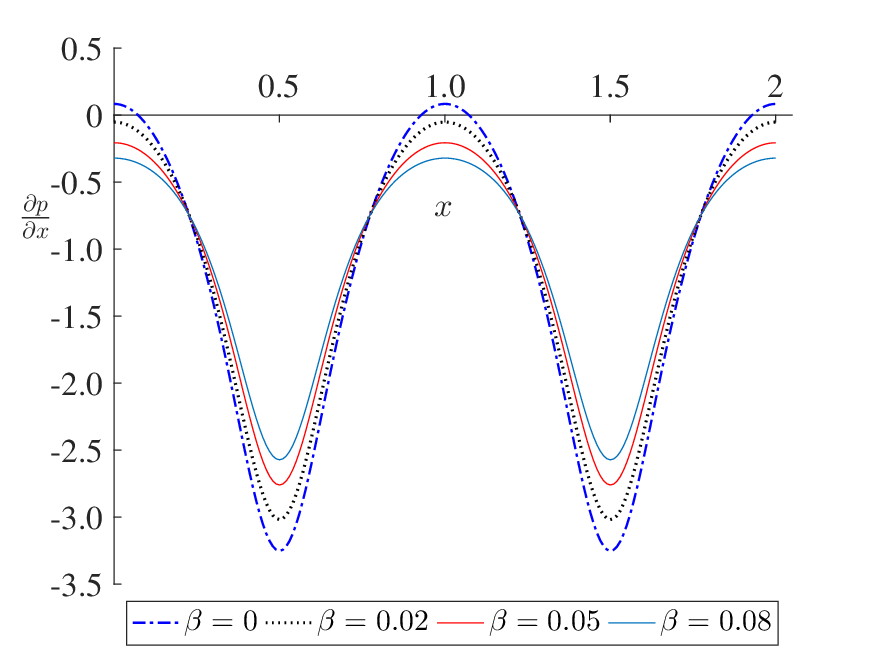}}
 \caption{{Effect of velocity slip on the wall pressure gradient for $\delta$ = 0.3, $a$ = 0.3, $\lambda=1.5$ and  (a) $Da = 10^{-1}$, (b) $Da = 4 \times 10^{-2}$}.} \label{f3}
 \end{figure}

Figures \ref{fig:8a}, \ref{fig:8c}, \ref{fig:8d} and \ref{fig:8b} demonstrate how the pressure distribution varies throughout the wall as a function of $x$ and $y$. The pressure gradient determines which way the flow goes. These figures illustrate how the wall slip modifies the curvature of the pressure distribution curves. The presence of velocity slip reduces wall shear resistance, while increasing anisotropic ratio introduces direction-dependent permeability. These effects alter the balance between driving and resistance forces, leading to a redistribution and modification of the pressure gradient profile. A relatively low Darcy number causes an irregular pressure distribution across the wall. It is observed that in the presence of slip, the pressure gradient remains negative along the wall for $\lambda=1,1.25,1.5$. However, in the absence of slip, it is negative only for $\lambda=1,1.25$ indicating the formation of a favourable pressure gradient. Notably, the pressure gradient changes as the value of $\lambda$ increases. Additionally, for slip flow the corresponding pressure gradients are shown in figures \ref{fig:8c}, \ref{fig:8d} and \ref{fig:8b}. It is observed that for $\lambda  \geq 1.75$, the pressure gradient is at the onset of a sign change near the crest of the wavy wall, indicating the development of an adverse pressure gradient that causes the flow reversal. In all circumstances, a symmetric pressure distribution about $x=1$ is seen.  
The existence of a slip at the walls with a relatively small Darcy number causes the maximum and minimum values of the pressure gradient to move somewhat upstream, as seen in the figures. In all situations, both high and low-pressure gradient peaks occur at $\lambda=2$. Further, for a fixed wall slip, it is noticed that diminishing the value of the porosity parameter $Da$ gradually increases the magnitude of the pressure gradient. It is also noted that by smaller porosity parameter $Da$ not only the magnitude of the pressure gradient is affected but also a uniform wavy pattern in the graph of the pressure gradient is disturbed. Thus, a remarkable role of the porosity parameter $Da$ and slip parameter $\beta$ on the distribution of pressure gradient is observed. For a fixed wall slip, a decrease in the Darcy number enhances porous resistance, requiring a larger pressure gradient to sustain the flow. Moreover, the increased resistance damps the flow's response to periodic variations, leading to distortion of the otherwise uniform wavy pressure gradient pattern. \\
Figures \ref{fig:wpslip} and \ref{wpslip1} demonstrate the effects of slip parameter $\beta$ on pressure distribution at two distinct Darcy numbers (Figure \ref{fig:wpslip} with $Da = 10^{-1}$ and Figure \ref{wpslip1}with $Da = 4 \times 10^{-2}$). The increase in the slip velocity parameter reduces wall shear resistance, allowing the pressure field to act more effectively in the flow direction, which results in the development of a favourable pressure gradient. This weakens the conditions necessary for flow reversal, thereby preventing the occurrence of backflow.
 \subsection{Shear stress distribution:}
\begin{figure}[ht!]
 \centering
 \subfigure[]{\label{fig:9a}\includegraphics*[width=7cm]{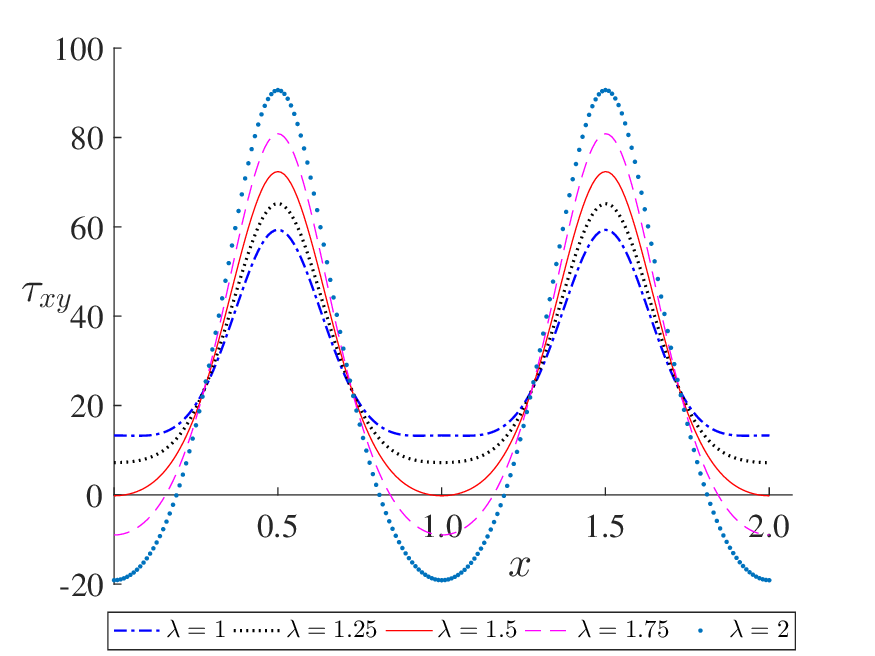}}\subfigure[]{\label{fig:9c}\includegraphics*[width=7cm]{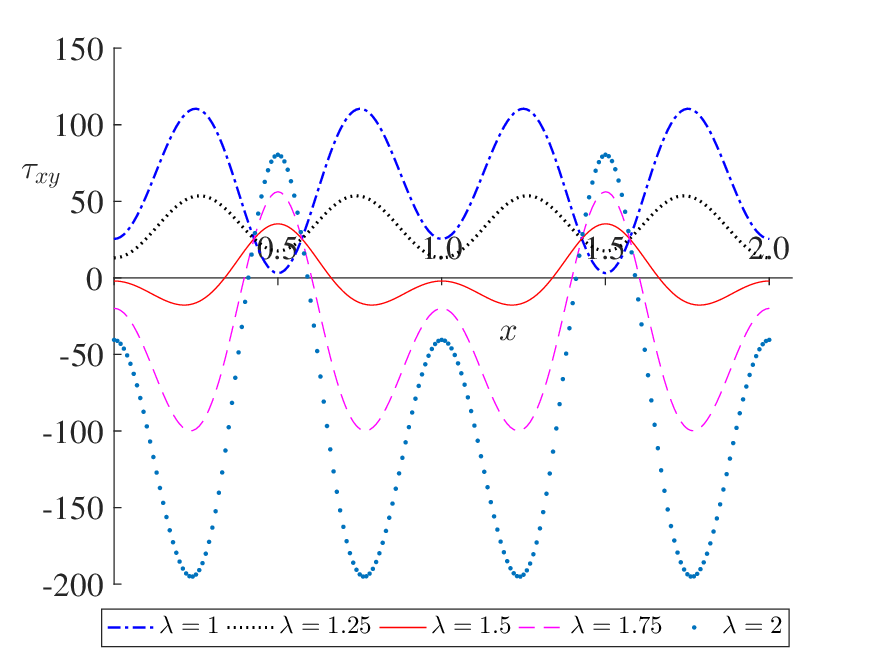}}
 \end{figure}

\begin{figure}[ht!]
 \centering
 \subfigure[]{\label{fig:9d}\includegraphics*[width=7cm]{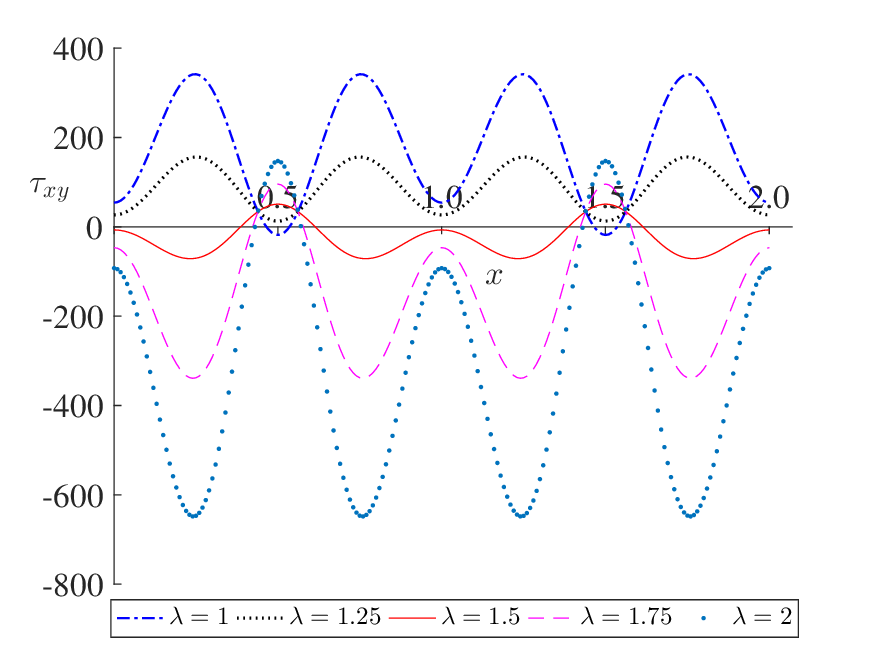}}\subfigure[]{\label{fig:9b}\includegraphics*[width=7cm]{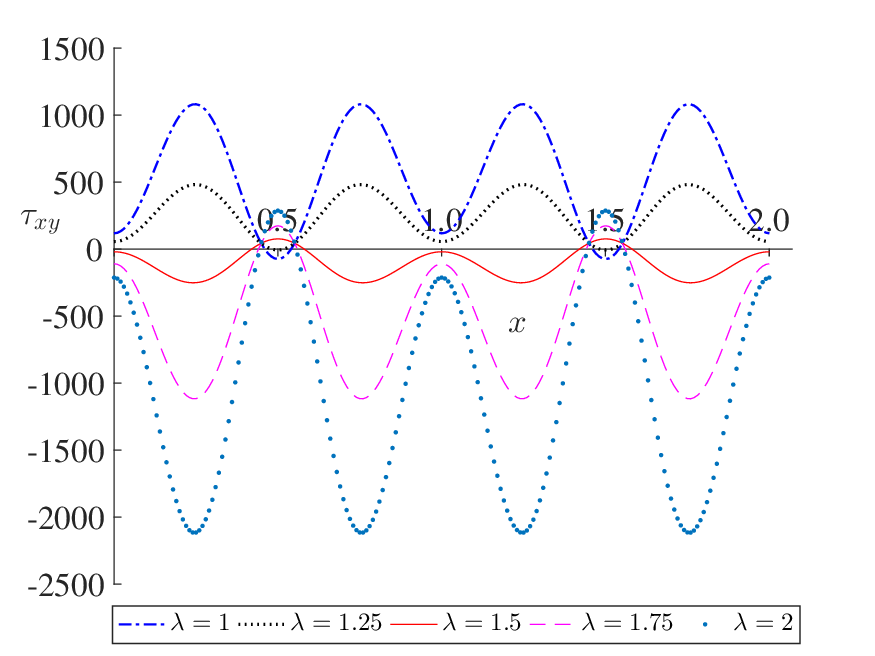}}
 \caption{Wall shear stress for $\delta$ = 0.3, $a$ = 0.3 and increasing values of $\lambda$ (a) without velocity slip ($\beta$ = 0.0) and $Da = 10^{-3}$ (b) with velocity slip ($\beta$ = 0.01) and $Da = 4 \times 10^{-3}$ (c) with velocity slip ($\beta$ = 0.01) and $Da = 2 \times 10^{-3}$ 
    (d) with velocity slip ($\beta$ = 0.01) and $Da = 10^{-3}$.}
 \label{f3}
 \end{figure}

\begin{figure}[ht!]
 \centering
 \subfigure[]{\label{fig:ssslip}\includegraphics*[width=7cm]{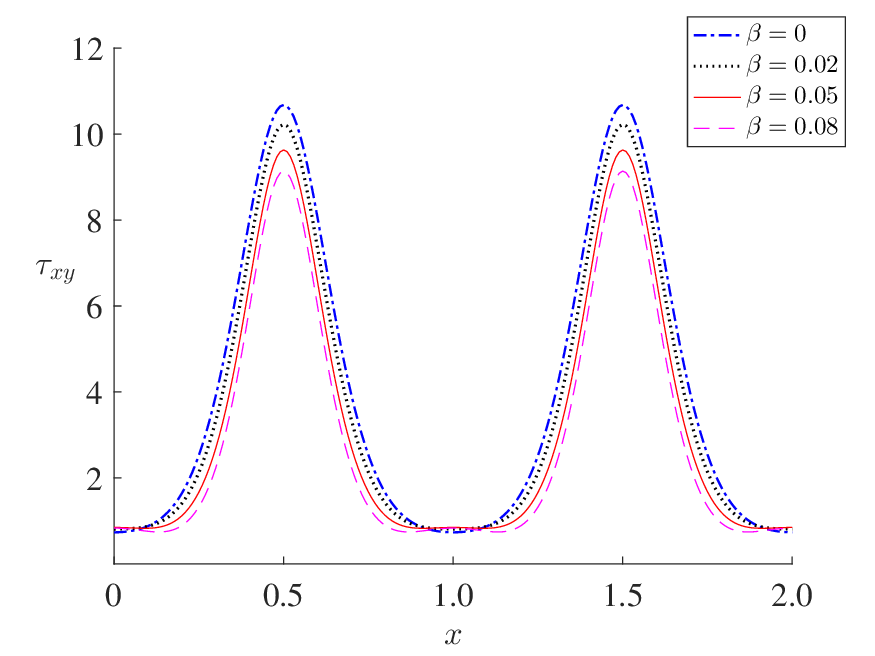}}\subfigure[]{\label{ssslip1}\includegraphics*[width=7cm]{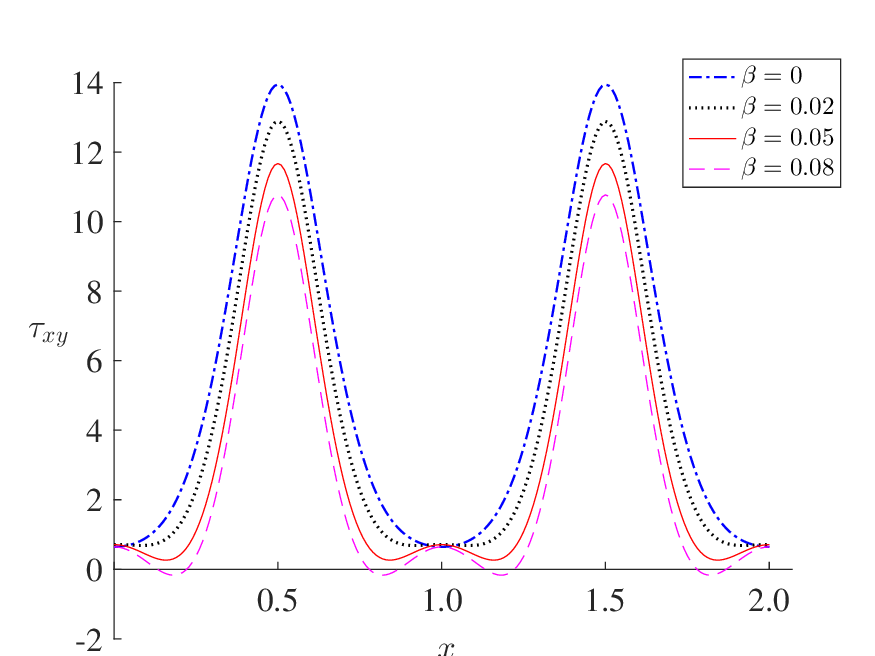}}
 \caption{{Effect of slip parameter $\beta$ on the wall shear stress for $\delta$ = 0.3, $a$ = 0.3, $\lambda=1.5$  and (a) $Da = 10^{-1}$, (b) $Da = 4 \times 10^{-2}$}.} \label{f3}
 \end{figure}

We now pay attention to the local shear stress distribution throughout the wall. Figures \ref{fig:9a}, \ref{fig:9c}, \ref{fig:9d} and \ref{fig:9b} illustrate the development of local shear stress at the top wall $y = B(x)$ for three different Darcy numbers ($Da = 10^{-3}, 2 \times 10^{-3}  4 \times 10^{-3} $). From those figures, it is observed that the existence of wall slip causes negative (favourable) shear stress corresponding to the high anisotropic ratio (for $\lambda = 1.5, 1.75, 2$), except at $x = 0.5, 1.5$, which pushes the primary direction of the motion. However, at a very small Darcy number value ($Da = 10^{-3}$), the existence of wall slip leads to a noticeable positive pressure distribution across the wall, corresponding to a small anisotropic ratio (for $\lambda = 1, 1.25$). With a fixed velocity slip parameter, the wall allows a controlled finite tangential velocity. When the anisotropic ratio increases, the permeability becomes highly direction-dependent, which strongly alters momentum transport in the flow. With fixed slip, increasing anisotropic ratio promotes flow reversal by enhancing resistance in the flow direction. From these figures it is observed that the existence of a slip at the wall alters somewhat the location of the sign change of wall shear stress to the right as compared to the no-slip scenario. It is also noticed that the wall shear stress distribution is symmetric about $x=1$. In all these subfigures, the irregularity in shear stress distribution across the wall associated with three different Darcy number values is clearly visualized.\\
  Figures \ref{fig:ssslip} and \ref{ssslip1} show how wall velocity slip affects the local shear stress distribution across the wall for two distinct Darcy number values (Figure \ref{fig:ssslip} with $Da = 10^{-1}$ and Figure \ref{ssslip1} with $Da = 4 \times 10^{-2}$) with $\lambda=1.5$. At a fixed anisotropic ratio, an increase in the velocity slip parameter reduces the near-wall velocity gradient, leading to a decrease in shear stress magnitude and the elimination of negative shear stress regions, thereby suppressing flow reversal. 
 
 \subsection{Axial velocity at the trough of the wavy wall:}
\begin{figure}[ht!]
 \centering
 \subfigure[]{\label{fig:10a}\includegraphics*[width=7cm]{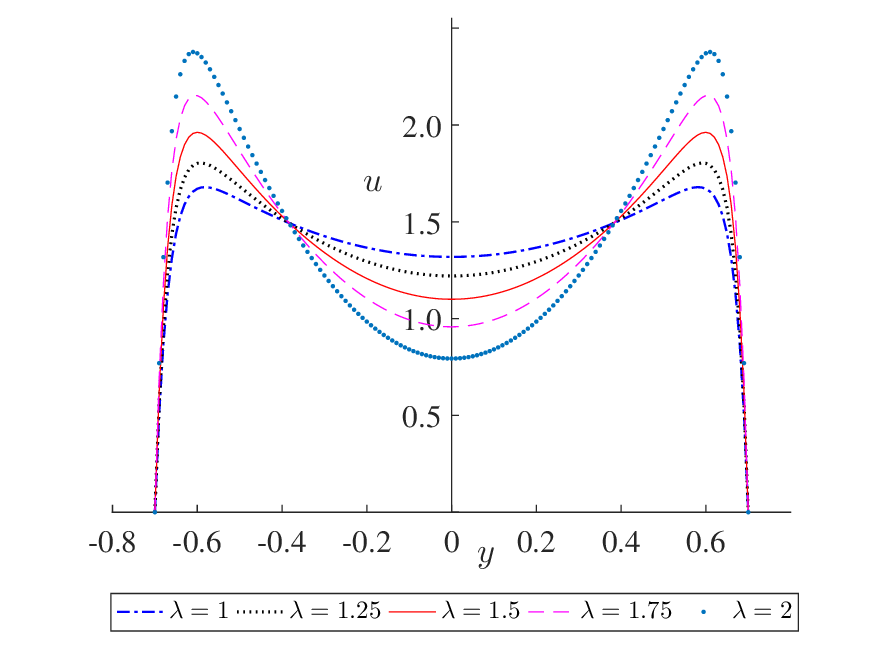}}\subfigure[]{\label{fig:10b}\includegraphics*[width=7cm]{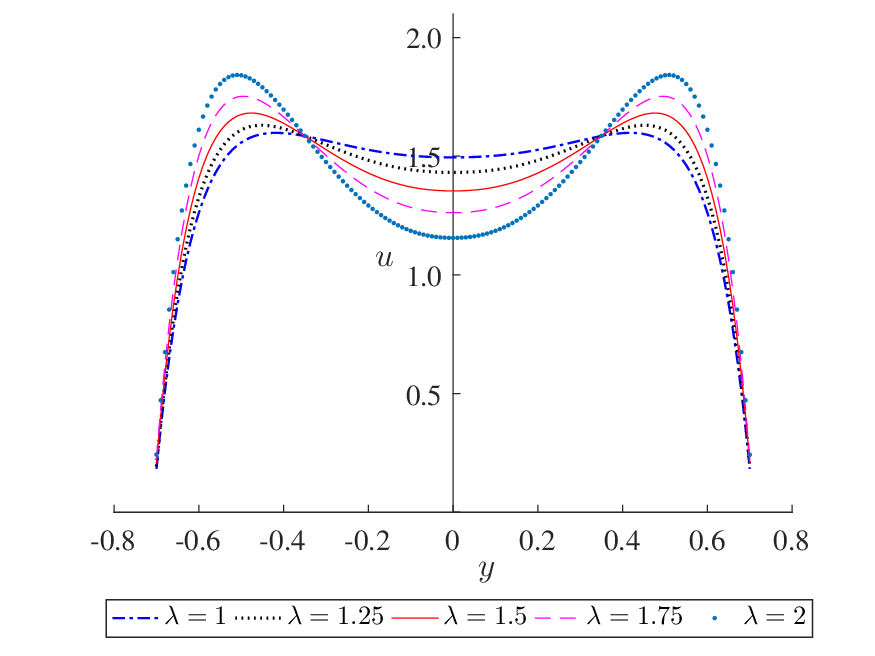}}
 \end{figure}

\begin{figure}[ht!]
 \centering
 \subfigure[]{\label{fig:10c}\includegraphics*[width=7cm]{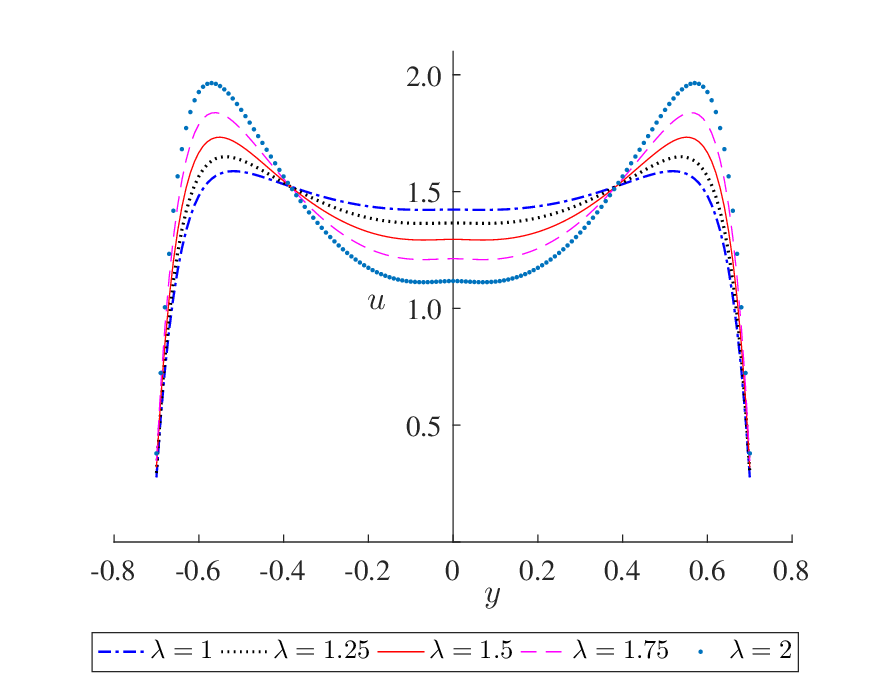}}\subfigure[]{\label{fig:10d}\includegraphics*[width=7cm]{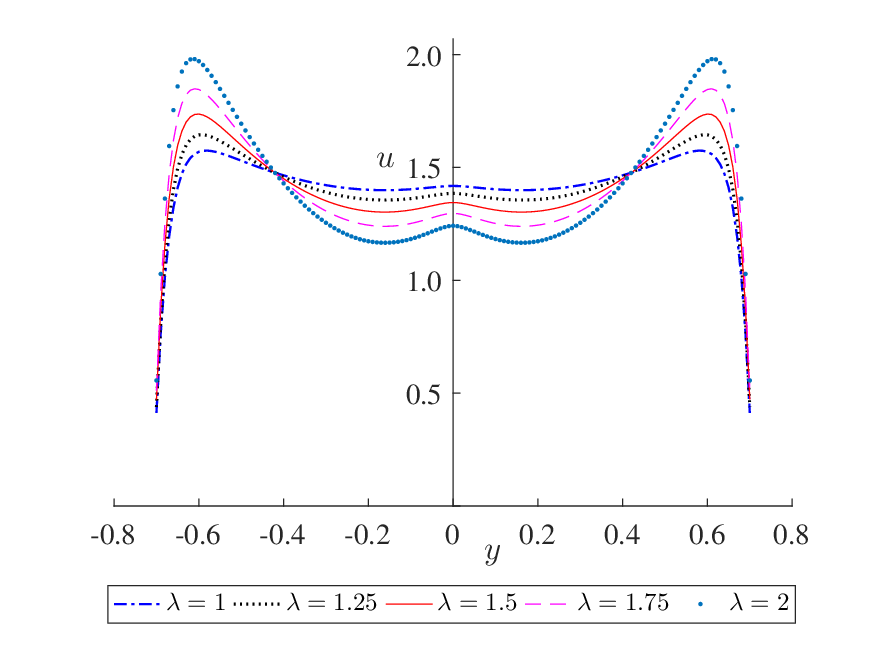}}
 \caption{ Impact of $\lambda$ on the velocity profile at the crest $(x = 0.5)$ for $\delta$ = 0.3, $a$ = 0.3, and (a) without velocity slip ($\beta = 0$) \& $Da = 10^{-3}$, (b) with velocity slip ($\beta$ = 0.01) \& $Da = 4 \times 10^{-3}$ (c) with velocity slip ($\beta$ = 0.01) \& $Da = 2 \times 10^{-3}$, (d) with velocity slip ($\beta$ = 0.01) \& $Da = 10^{-3}$.} \label{f3}
 \end{figure}

Figures \ref{fig:10a}, \ref{fig:10b}, \ref{fig:10c}, and \ref{fig:10d} illustrate the effect of $\lambda$ on the axial velocity at the through of the wavy wall for three different porosity parameter values ($Da=10^{-3}, 4 \times 10^{-3}, 2 \times 10^{-3}$). The results are presented for both cases: without velocity slip (Fig. \ref{fig:10a}, where $\beta=0$) and with velocity slip (Figs. \ref{fig:10b}, \ref{fig:10c}, and \ref{fig:10d}, where $\beta=0.01$) at the wavy walls.
In addition, Figures \ref{fig:3a}, \ref{fig:3b}, \ref{fig:3c}, \ref{fig:3d} and figures \ref{fig:10a}, \ref{fig:10b}, \ref{fig:10c}, \ref{fig:10d}  indicate that the deformation of the velocity at crest is reversed from the axial velocity at the trough.
From those sub-figures, it is observed that velocity near the wall is higher than in the middle. The axial pressure gradient at the wavy wall can provide a detailed explanation of this phenomena. If we pay close attention to Figures \ref{fig:8a}, \ref{fig:8c}, \ref{fig:8d}, \ref{fig:8b}, we can observe an adverse pressure gradient being generated near the crest. However, a negative pressure gradient directs the flow in the trough, resulting in high velocity near the wavy wall and low velocity toward the middle of the wavy channel to keep the volumetric flow rate constant. With fixed velocity slip ($\beta=0.01$), increasing anisotropic ratio induces directional resistance that redistributes momentum across the channel. Due to reduced wall damping, this redistribution manifests as an abrupt variation in the velocity profile near the centerline. Additionally, decreases in the porosity parameter $Da$ modify the curvature of the velocity profile.
     
\begin{figure}[t]
 \centering
 \subfigure[]{\label{fig:11a}\includegraphics*[width=7cm]{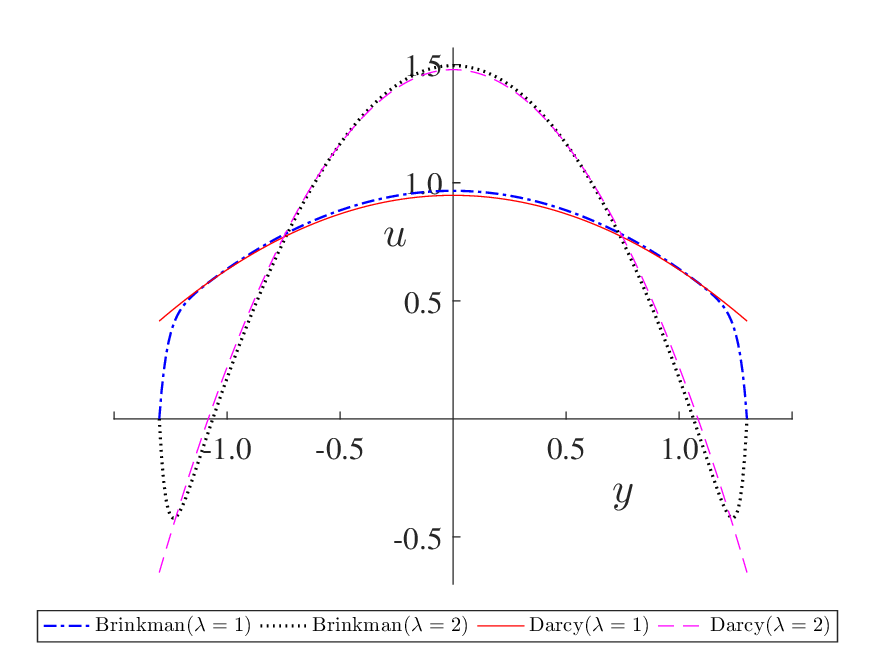}}\subfigure[]{\label{fig:11b}\includegraphics*[width=7cm]{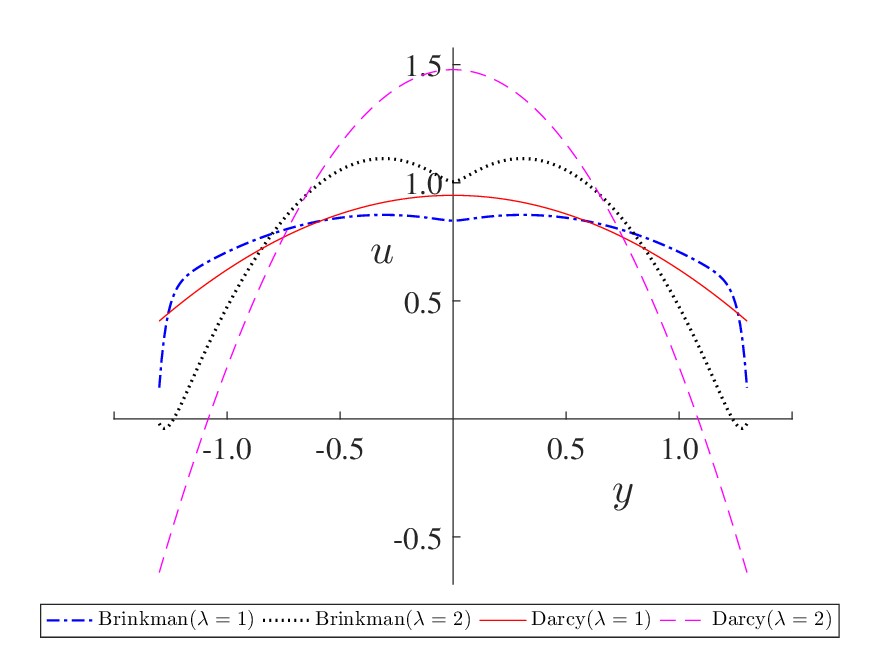}}
 \caption{The effect of $\lambda$ on the velocity profile at the crest (x = 0) for $\delta$ = 0.3, $a$ = 0.8, $Da = 10^{-3}$ and (a) without velocity slip ($\beta$ = 0.0), (b) with velocity slip ($\beta = 0.01).$}\label{f3}
 \end{figure}

\begin{figure}[t]
 \centering
 \subfigure[]{\label{fig:12a}\includegraphics*[width=7cm]{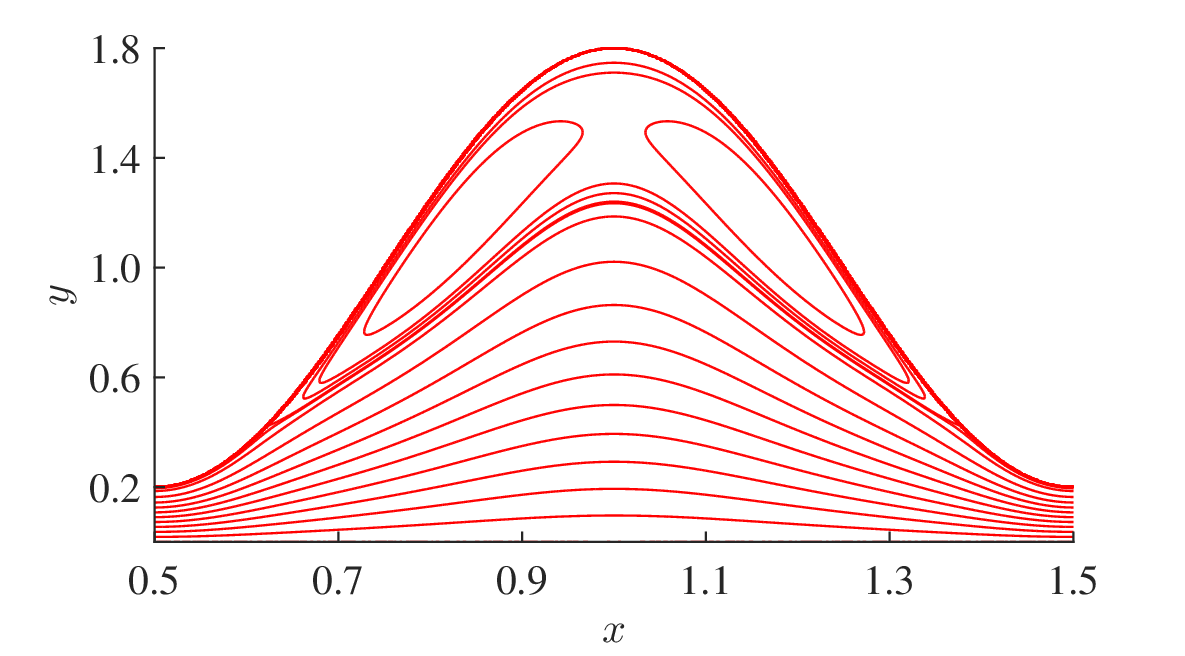}}\subfigure[]{\label{fig:12bb}\includegraphics*[width=7.9cm]{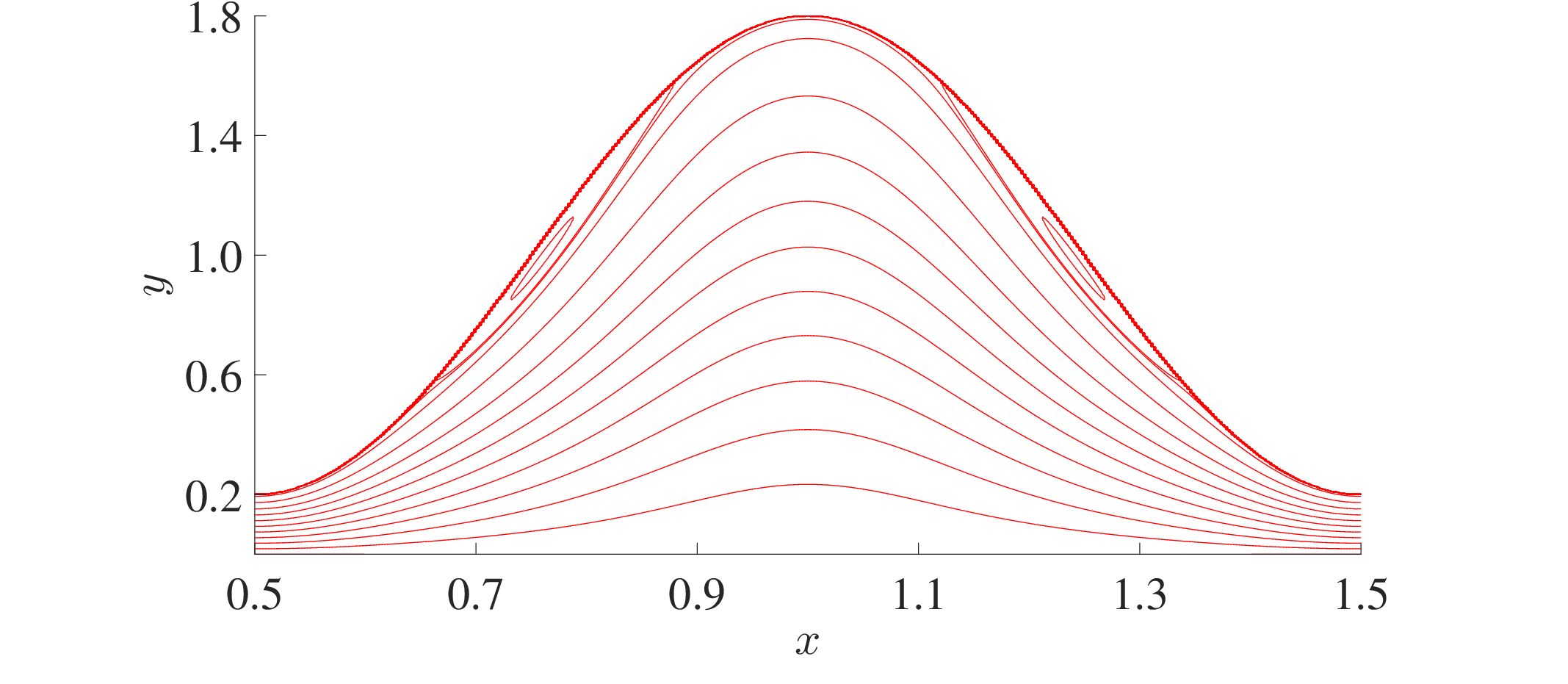}}
 \caption{ Enlarged view of streamline near crest for $\delta$ = 0.3, $a$ = 0.8, $Da = 10^{-3}$, $\lambda = 1$  and (a) without velocity slip ($\beta$ = 0.0), (b) with velocity slip ($\beta = 0.03).$}\label{f3}
 \end{figure}

Figures \ref{fig:11a} and \ref{fig:11b} show the comparison of velocity between the Darcy and the Brinkman equations at the crest. Figures \ref{fig:12a} and \ref{fig:12bb} present close-up views of the streamline variations depicted in Figures \ref{Threshold_fig 16} and \ref{Threshold_fig 18}. These variations are shown within the region $0.5\leq x \leq 1.5 $ in an isotropic porous medium, corresponding to $\lambda=1$ and for a high amplitude parameter $(a=0.8)$. Two co-rotating vortices at the crest of the wavy wall are seen in figure \ref{fig:12a}. In other words, if the amplitude parameter is big enough, flow separation may occur in an isotropic porous material. Figure \ref{fig:12bb} shows that, in such a situation also the flow reversal may be prevented by selecting a proper slip value at the wall. In an isotropic porous medium, large wall amplitude induces strong adverse pressure gradients leading to flow separation. However, the introduction of wall slip reduces viscous resistance and enhances near-wall momentum, thereby weakening the adverse pressure effects and suppressing flow separation. Figure \ref{fig:12bb} further shows that the existence of slip at the wall causes the streamline near $x = 1$ to deviate away from the channel centerline.

\section{Reverse flow region}\label{sec6}
The flow over the 2D and 3D wavy walls is characterized by flow reversal phenomena behind the wave crest. According to the computations stated in section \ref{sec4}, one can represent the axial velocity as
\begin{eqnarray}
u \sim u_0+\delta^2 u_1,\\ 
u_0 \sim - p_{0x} ,\\
p_{0x} \sim -\frac{1}{h(x)},\\
u_1 \sim \lambda^2 \frac{p_{0xxx}}{2} y^2 +d_1(x) ,\\
d_1(x) \sim -\lambda^2 \frac{p_{0xxx}}{6} h(x)^2,
\end{eqnarray}
and 
\begin{eqnarray} \label{60}
u \sim \frac{1}{h(x)}- \frac{\delta^2 \lambda^2 }{2}\left(y^2-\frac{h(x)^2}{3}\right) \frac{d^2}{dx^2} \left(\frac{1}{h(x)}\right).
\end{eqnarray}
It is noted from the equation \eqref{60}, that for the slip flow ($\beta \ne 0$) 
with $\alpha \gg 1$, the axial velocity is dependent on numerous characteristics, such as the anisotropic ratio $\lambda$, aspect ratio $\delta$, amplitude parameter $a$ and so on. We highlight the circulation zone under the crest $(x=0)$ of the wavy wall. To do this, let's look at the velocity at $x=0$, which is provided by \\
\begin{eqnarray} \label{80}
u(x=0, y) \sim \frac{1}{1+a}- \frac{ 2 \delta^2 \lambda^2 a \pi^2}{(1+a)^2}\left(y^2-\frac{(1+a)^2}{3} \right).
\end{eqnarray}
The circulation happens near the crest ($x = 0$) of the wavy wall when the velocity turns negative, i.e.\\
\begin{eqnarray} \label{90}
u(x=0, y) < 0,
\end{eqnarray}
whenever
\begin{eqnarray} \label{50}
|y| > \sqrt{\frac{(1+a)^2}{3}+ \frac{(1+a)}{2 \delta^2 \lambda^2 a \pi^2}}.
\end{eqnarray}
Equation \eqref{50} indicates the range of $y$ for the flow separation zone near the top of the wavy wall for the slip flow ($\beta \neq 0$) 
with $\alpha \gg 1$. Thus the physical influence of the slip parameters $\beta$ on the flow reversal phenomena is clearly visible.  We've shown earlier how this flow separation can be managed by appropriate slip parameters. Karmakar et al. \cite{karmakar2017note} investigated the corresponding flow reversal zone for the case of zero velocity at the wavy walls. It may be recovered from the current study by taking $\beta = 0$ in the equation \eqref{70}. Then for $\alpha \gg 1$, we obtain from the equation \eqref{70}, $Q(x) \sim -h(x)$ as $|\tanh(\alpha h(x))|\leq 1$ and from the equation \eqref{60} one may obtain for  $\alpha \gg 1$,\\
\begin{eqnarray}
u(x, y) \sim \frac{1}{h(x)}- \frac{\delta^2 \lambda^2 }{2}\left(y^2-\frac{h(x)^2}{3}\right) \frac{d^2}{dx^2} \left(\frac{1}{h(x)}\right).
\end{eqnarray}
For no-slip flow (i.e. $\beta=0$), the range for the flow reversal zone for $\alpha \gg 1$  may be calculated with the help of \eqref{80}, \eqref{90} and \eqref{50}. Thus the existing results of Karmakar et al.  \cite{karmakar2017note} regarding flow reversal phenomena may be recovered by our study in a limiting case $\beta=0$.

\section{Limiting cases}\label{sec7}
In this section, we describe the unperturbed flow situation by considering $a \sim 0$, which indicates $h(x) \sim 1$. Consequently, 
\begin{eqnarray}
a_1(x) = \frac{p_{0x}}{\cosh(\alpha )-\beta\alpha\sinh(\alpha )},
\end{eqnarray}
\begin{eqnarray}
			a_2(x) = 0,
\end{eqnarray}
\begin{eqnarray}
p_{0x} = \frac{\alpha[\cosh(\alpha )-\beta\alpha\sinh(\alpha )]}{\sinh(\alpha )-\alpha[\cosh(\alpha )-\beta\alpha \sinh(\alpha )]},
\end{eqnarray}
\begin{eqnarray} \label{150}
u_0 \sim \frac{\alpha[\cosh(\alpha y)-\cosh(\alpha)+\alpha \beta \sinh(\alpha)]}{\sinh(\alpha)-\alpha[\cosh(\alpha)-\alpha \beta \sinh(\alpha)]}.
\end{eqnarray}
It is also observed that in the situation of unperturbed flow (i.e. $\delta = 0, 
a = 0$), if the permeability along the $x$-direction is very high then $\alpha \ll 1$, and
\begin{eqnarray} \label{200}
\cosh(\alpha) \sim 1+\frac{\alpha^2}{2!}+O(\alpha^4),
\end{eqnarray}
\begin{eqnarray} \label{250}
\sinh(\alpha) \sim \alpha+\frac{\alpha^3}{3!}+O(\alpha^5).
\end{eqnarray}
Consequently, using \eqref{200} and \eqref{250} in \eqref{150}, one may get the expression for $u_0(y)$ as, \\
\begin{eqnarray}
u_0(y) \sim \frac{\alpha[1+\frac{\alpha^2 y^2}{2!}-1-\frac{\alpha^2}{2!}+ \alpha \beta (\alpha+\frac{\alpha^3}{3!})]}{\alpha+\frac{\alpha^3}{3!}-\alpha[1+\frac{\alpha^2}{2!}-\alpha \beta(\alpha+ \frac{\alpha^3}{3!})]},
\end{eqnarray}
i.e., 
\begin{eqnarray}
u_0(y) \sim \frac{\frac{y^2}{2}-\frac{1}{2}+\beta+\frac{\alpha^2\beta}{6}}{-\frac{1}{3}+\beta+\frac{\alpha^2\beta}{6}}.
\end{eqnarray}
Now $\alpha\ll 1$  and $\beta$ is very small slip coefficient $\implies$ $\frac{\alpha^2\beta}{6} \sim 0$\\
i.e, 
			
\begin{eqnarray} \label{100}
u_0(y) \sim \frac{\frac{3}{2}(1-y^2)-3\beta}{1-3 \beta}.
\end{eqnarray} 
For a fixed value of the slip parameter $\beta$, equation \eqref{100} represents the velocity related to the plane Poiseuille flow. 
		
\section{Summary and conclusion}\label{sec8}
Throughout this study, an extensive analysis of the flow dynamics in a symmetric channel with wavy walls is conducted by considering the influence of wall velocity slip. The study considers an anisotropic porous medium, which plays a crucial role in influencing the flow characteristics. In this study, we have used the Darcy-Brinkman equation as momentum conservation equation. The impermeable walls are taken that allow tangential velocity slip, and the flow velocity at the wavy walls satisfies the Navier slip boundary criteria. The walls of the channel are coated with the same materials, and the case is handled by imposing symmetric Navier slip conditions. This study is the extension of the work of Karmakar et al. \cite{karmakar2017note} by incorporating velocity slips at the wavy walls. The current investigation demonstrates a strong agreement with established solutions in limiting cases from the existing literature and validating their accuracy. The axial velocity profile is initially obtained for plane Poiseuille flow by considering the limiting values of the flow parameters and validating the results against established findings in the existing literature \textcolor{red}{[Put Ref.]}. The principal motivation of this work is to report how the parameters $\lambda$, $\beta$ and $Da$ influence the solution of the flow system. We have solved the flow fields using the perturbation method for a small aspect ratio $\delta$, up to order $O(\delta^2)$. The flow dynamics are explored by addressing the graphical representations of axial velocity profiles, streamline variations, pressure distributions, and shear stress behavior. All these flow features are mainly characterized by the dimensionless parameters $a$, $\delta$, $\lambda$, $Da$, and $\beta$. Streamlines, axial velocity, and all other hydrodynamic quantities of the underlying symmetric flow are regulated by the modeled parameters ($\delta$, $a$, $Da$, and $\lambda$) together with the slip parameter $\beta$.
The obtained results provide deeper insights into the impact of slip conditions and porous anisotropy on the velocity profiles and overall fluid behavior. The Darcy–Brinkman model extends classical Darcy flow by including viscous shear effects (through a Laplacian term), which allows it to describe boundary layers and more complex flow structures—like vortices—especially near solid walls such as a wavy surface. Introduction of velocity slip at the wall  effectively allowing the fluid to move more freely along the boundary. As the slip parameter increases, the shear stress at the wall decreases, the velocity gradient near the wall becomes weaker and flow separation is suppressed. Vortices near the crest of a wavy wall are typically generated by strong shear and adverse pressure gradients. With stronger slip the fluid doesn’t “stick” and slow down as much at the wall. This reduces the buildup of shear layers that would otherwise roll up into vortices.
As a result, vortex formation is weakened or even eliminated. A wavy channel introduces spatially varying geometry, leading to flow separation. Slip acts as a flow smoother in geometrically complex domains. It reduces energy dissipation near walls.
It mitigates geometric-induced disturbances. Flow reversal in wavy-channel Darcy–Brinkman equation systems is a rich interplay between geometry, inertia, and porous resistance. The key idea is that even when the net pressure gradient drives fluid forward, local regions can experience backward flow due to separation and recirculation induced by wall waviness and porous drag. The study reveals that, the slip parameter $\beta$ as well as the porosity parameter $Da$ controls the curvature of the velocity profile. Moreover, the existence of slip at the walls reduces the backflow phenomena. The overall pressure gradient is also altered under the influence of the slip parameter. It is observed that the stronger velocity slip reduces the number of vortices near the crest of the wavy wall and in the isotropic porous medium with a large amplitude parameter, the flow reversal may also be controlled. Conclusively, to control the flow characteristics passively as per the requirement of applications, the channel wall can be designed with suitable materials that permit partial velocity slip.

In summary, the study highlights that wall velocity slip, anisotropic ratio, and porosity parameters play a crucial role in shaping flow behavior in confined geometries. Flow reversal in symmetric flow can be mitigated by designing boundary walls with hydrophobic, slippery, coated, or porous surfaces with low permeability. The Navier slip condition, with an appropriate slip length, provides a reliable approximation for wall velocity. Such kind of study has significant applications across various fields, from lubrication to microfluidics, making it valuable for numerous real-world technologies.

\section*{Funding}{First author acknowledges the financial support provided by the Council of Scientific \& Industrial Research, India, through the CSIR-UGC NET fellowship for pursuing the doctoral degree.}

\section*{Declaration of interests}{The authors declare that they have no conflict of interest.}

\section*{Data availability statement}{All the data that support the findings of this investigation are available within the article.}

\section*{Author contributions}{ First author: Methodology, Formal analysis, Writing—original draft, and software.
Second author: Formal analysis, Writing, review, editing, and supervision.
Third author: Formal analysis, Investigation, Software.
Fourth author: Conceptualization, Investigation, Methodology, Formal analysis, Writing—original draft, and software.}


       
     

	\bibliography{0ref}

@article{DA,
  title={Thermally developing forced convection in a porous medium: parallel plate channel with walls at uniform temperature, with axial conduction and viscous dissipation effects},
  author={Nield, DA and Kuznetsov, AV and Xiong, Ming},
  journal={International Journal of Heat and Mass Transfer},
  volume={46},
  number={4},
  pages={643--651},
  year={2003},
  publisher={Elsevier}
}

@article{K,
  title={Effects of viscous dissipation on thermally developing forced convection in a porous saturated circular tube with an isoflux wall},
  author={Hooman, Kamel and Pourshaghaghy, Alireza and Ejlali, Arash},
  journal={Applied Mathematics and Mechanics},
  volume={27},
  pages={617--626},
  year={2006},
  publisher={Springer}
}

@article{karmakar2016effect,
  title={Effect of anisotropic permeability on fluid flow through composite porous channel},
  author={Karmakar, Timir and Raja Sekhar, GP},
  journal={Journal of Engineering Mathematics},
  volume={100},
  number={1},
  pages={33--51},
  year={2016},
  publisher={Springer}
}

@book{nield2006convection,
  title={Convection in porous media},
  author={Nield, Donald A and Bejan, Adrian and others},
  volume={3},
  year={2006},
  publisher={Springer}
}

@article{gray2013darcy,
  title={Darcy flow in a wavy channel filled with a porous medium},
  author={Gray, Donald D and Ogretim, Egemen and Bromhal, Grant S},
  journal={Transport in porous media},
  volume={98},
  number={3},
  pages={743--753},
  year={2013},
  publisher={Springer}
}

@article{datta2010study,
  title={Study of steady viscous flow through a wavy channel: non-orthogonal coordinates},
  author={Datta, Sunil and Tripathi, Akhilesh},
  journal={Meccanica},
  volume={45},
  pages={809--815},
  year={2010},
  publisher={Springer}
}

@article{huitt1956fluid,
  title={Fluid flow in simulated fractures},
  author={Huitt, JL},
  journal={AIChE Journal},
  volume={2},
  number={2},
  pages={259--264},
  year={1956},
  publisher={Wiley Online Library}
}

@article{ng2010darcy,
  title={Darcy--Brinkman flow through a corrugated channel},
  author={Ng, Chiu-On and Wang, CY},
  journal={Transport in porous media},
  volume={85},
  pages={605--618},
  year={2010},
  publisher={Springer}
}

@article{karmakar2021physics,
  title={Physics of unsteady Couette flow in an anisotropic porous medium},
  author={Karmakar, Timir},
  journal={Journal of Engineering Mathematics},
  volume={130},
  number={1},
  pages={8},
  year={2021},
  publisher={Springer}
}

@article{rice1970anisotropic,
  title={Anisotropic permeability in porous media},
  author={Rice, Philip A and Fontugne, Daniel J and Latini, Raimondo G and Barduhn, Allen J},
  journal={Industrial \& Engineering Chemistry},
  volume={62},
  number={6},
  pages={23--31},
  year={1970},
  publisher={ACS Publications}
}

@article{kumar2020elastohydrodynamics,
  title={Elastohydrodynamics of a deformable porous packing in a channel competing under shear and pressure gradient},
  author={Kumar, Prakash and Sekhar, GP},
  journal={Physics of Fluids},
  volume={32},
  number={6},
  year={2020},
  publisher={AIP Publishing}
}

@article{brinkman1949calculation,
  title={A calculation of the viscous force exerted by a flowing fluid on a dense swarm of particles},
  author={Brinkman, Hendrik C},
  journal={Flow, Turbulence and Combustion},
  volume={1},
  number={1},
  pages={27--34},
  year={1949},
  publisher={Springer}
}

@article{nield1983boundary,
  title={The boundary correction for the Rayleigh-Darcy problem: limitations of the Brinkman equation},
  author={Nield, D Ao},
  journal={Journal of Fluid Mechanics},
  volume={128},
  pages={37--46},
  year={1983},
  publisher={Cambridge University Press}
}

@article{karmakar2017note,
  title={A note on flow reversal in a wavy channel filled with anisotropic porous material},
  author={Karmakar, Timir and Raja Sekhar, GP},
  journal={Proceedings of the Royal Society A: Mathematical, Physical and Engineering Sciences},
  volume={473},
  number={2203},
  pages={20170193},
  year={2017},
  publisher={The Royal Society Publishing}
}

@book{navier1822memoire,
  title={M{\'e}moire sur les lois du mouvement des fluides},
  author={Navier, Claude},
  year={1822},
  publisher={{\'e}diteur inconnu}
}

@book{helmholtz1860reibung,
  title={{\"U}ber Reibung tropfbarer Fl{\"u}ssigkeiten: Von H. Helmholtz und G. v. Piotrowski.(Mit 2 Taff.)(Aus d. XL. Bd. S. 607. 1860. der Sitzgsber. der math-nat. Cl. der k. Ak. der Wiss. bes. dbg.)},
  author={Helmholtz, H},
  year={1860},
  publisher={Hof-\& Stts.-Druck}
}

@article{rao1999effect,
  title={The effect of the slip boundary condition on the flow of fluids in a channel},
  author={Rao, IJ and Rajagopal, KR},
  journal={Acta Mechanica},
  volume={135},
  number={3-4},
  pages={113--126},
  year={1999},
  publisher={Springer}
}

@article{migler1993slip,
  title={Slip transition of a polymer melt under shear stress},
  author={Migler, KB and Hervet, H and Leger, L},
  journal={Physical review letters},
  volume={70},
  number={3},
  pages={287},
  year={1993},
  publisher={APS}
}

@article{migler1994slip,
  title={The slip transition at the polymer-solid interface},
  author={Migler, KB and Massey, G and Hervet, I and Leger, L},
  journal={Journal of Physics: Condensed Matter},
  volume={6},
  number={23A},
  pages={A301},
  year={1994},
  publisher={IOP Publishing}
}

@article{min2005effects,
  title={Effects of hydrophobic surface on stability and transition},
  author={Min, Taegee and Kim, John},
  journal={Physics of fluids},
  volume={17},
  number={10},
  year={2005},
  publisher={AIP Publishing}
}

@article{chattopadhyay2016yih,
  title={On the Yih--Marangoni instability of a two-phase plane Poiseuille flow in a hydrophobic channel},
  author={Chattopadhyay, Geetanjali and Usha, R},
  journal={Chemical Engineering Science},
  volume={145},
  pages={214--232},
  year={2016},
  publisher={Elsevier}
}

@article{ghosh2014linear,
  title={Linear stability analysis of miscible two-fluid flow in a channel with velocity slip at the walls},
  author={Ghosh, Sukhendu and Usha, R and Sahu, Kirti Chandra},
  journal={Physics of Fluids},
  volume={26},
  number={1},
  year={2014},
  publisher={AIP Publishing}
}

@article{karmakar2016lifting,
  title={Lifting a large object from an anisotropic porous bed},
  author={Karmakar, Timir and Raja Sekhar, GP},
  journal={Physics of Fluids},
  volume={28},
  number={9},
  year={2016},
  publisher={AIP Publishing}
}

@article{degan2002forced,
  title={Forced convection in horizontal porous channels with hydrodynamic anisotropy},
  author={Degan, G and Zohoun, S and Vasseur, P},
  journal={International Journal of Heat and Mass Transfer},
  volume={45},
  number={15},
  pages={3181--3188},
  year={2002},
  publisher={Elsevier}
}

@article{wei2003flow,
  title={Flow in a wavy-walled channel lined with a poroelastic layer},
  author={Wei, HH and Waters, SL and Liu, Shu Qian and Grotberg, JB},
  journal={Journal of Fluid Mechanics},
  volume={492},
  pages={23--45},
  year={2003},
  publisher={Cambridge University Press}
}

@article{tsangaris1984laminar,
  title={On laminar steady flow in sinusoidal channels},
  author={Tsangaris, S and Leiter, E},
  journal={Journal of engineering mathematics},
  volume={18},
  number={2},
  pages={89--103},
  year={1984},
  publisher={Springer}
}

@article{luo2008two,
  title={Two-layer flow in a corrugated channel},
  author={Luo, H and Blyth, MG and Pozrikidis, C},
  journal={Journal of Engineering Mathematics},
  volume={60},
  pages={127--147},
  year={2008},
  publisher={Springer}
}

@article{taylor1971model,
  title={A model for the boundary condition of a porous material. Part 1},
  author={Taylor, GI t},
  journal={Journal of Fluid Mechanics},
  volume={49},
  number={2},
  pages={319--326},
  year={1971},
  publisher={Cambridge University Press}
}

@article{hill2008poiseuille,
  title={Poiseuille flow in a fluid overlying a porous medium},
  author={Hill, Antony A and Straughan, Brian},
  journal={Journal of Fluid Mechanics},
  volume={603},
  pages={137--149},
  year={2008},
  publisher={Cambridge University Press}
}

@article{kitanidis1997stokes,
  title={Stokes flow in a slowly varying two-dimensional periodic pore},
  author={Kitanidis, Peter K and Dykaar, Bruce B},
  journal={Transport in porous media},
  volume={26},
  pages={89--98},
  year={1997},
  publisher={Springer}
}

@article{malevich2006stokes,
  title={Stokes flow through a channel with wavy walls},
  author={Malevich, AE and Mityushev, VV and Adler, PM},
  journal={Acta mechanica},
  volume={182},
  pages={151--182},
  year={2006},
  publisher={Springer}
}

@article{yu2013darcy,
  title={Darcy-Brinkman flow through a bumpy channel},
  author={Yu, LH and Wang, CY},
  journal={Transport in porous media},
  volume={97},
  number={3},
  pages={281--294},
  year={2013},
  publisher={Springer}
}

@article{grattoni1995anisotropy,
  title={Anisotropy in pore structure of porous media},
  author={Grattoni, Carlos A and Dawe, Richard A},
  journal={Powder technology},
  volume={85},
  number={2},
  pages={143--151},
  year={1995},
  publisher={Elsevier}
}

@article{yovogan2013effect,
  title={Effect of anisotropic permeability on convective heat transfer through a porous river bed underlying a fluid layer},
  author={Yovogan, Julien and Degan, G{\'e}rard},
  journal={Journal of Engineering Mathematics},
  volume={81},
  number={1},
  pages={127--140},
  year={2013},
  publisher={Springer}
}

@article{haddad2007forced,
  title={Forced convection of gaseous slip-flow in porous micro-channels under Local Thermal Non-Equilibrium conditions},
  author={Haddad, OM and Al-Nimr, MA and Al-Omary, J Sh},
  journal={Transport in porous media},
  volume={67},
  pages={453--471},
  year={2007},
  publisher={Springer}
}

@article{javid2022porosity,
  title={Porosity effects on the peristaltic flow of biological fluid in a complex wavy channel},
  author={Javid, Khurram and Asghar, Zeeshan and Saeed, Umer and Waqas, Muhammad},
  journal={Pramana},
  volume={96},
  number={1},
  pages={2},
  year={2022},
  publisher={Springer}
}

@article{tripathi2013study,
  title={Study of transient peristaltic heat flow through a finite porous channel},
  author={Tripathi, Dharmendra},
  journal={Mathematical and Computer Modelling},
  volume={57},
  number={5-6},
  pages={1270--1283},
  year={2013},
  publisher={Elsevier}
}

@article{zeeshan2018convective,
  title={Convective Poiseuille flow of Al2O3-EG nanofluid in a porous wavy channel with thermal radiation},
  author={Zeeshan, Ahmad and Shehzad, Nasir and Ellahi, Rahmat and Alamri, Sultan Z},
  journal={Neural Computing and Applications},
  volume={30},
  number={11},
  pages={3371--3382},
  year={2018},
  publisher={Springer}
}

@article{sharipov2004velocity,
  title={Velocity slip and temperature jump coefficients for gaseous mixtures. II. Thermal slip coefficient},
  author={Sharipov, Felix and Kalempa, Denize},
  journal={Physics of Fluids},
  volume={16},
  number={3},
  pages={759--764},
  year={2004},
  publisher={American Institute of Physics}
}

@article{pit2000direct,
  title={Direct experimental evidence of slip in hexadecane: solid interfaces},
  author={Pit, R and Hervet, H and Leger, L},
  journal={Physical review letters},
  volume={85},
  number={5},
  pages={980},
  year={2000},
  publisher={APS}
}

@article{bonaccurso2002hydrodynamic,
  title={Hydrodynamic force measurements: boundary slip of water on hydrophilic surfaces and electrokinetic effects},
  author={Bonaccurso, Elmar and Kappl, Michael and Butt, Hans-J{\"u}rgen},
  journal={Physical Review Letters},
  volume={88},
  number={7},
  pages={076103},
  year={2002},
  publisher={APS}
}

@article{zhu2002limits,
  title={Limits of the hydrodynamic no-slip boundary condition},
  author={Zhu, Yingxi and Granick, Steve},
  journal={Physical review letters},
  volume={88},
  number={10},
  pages={106102},
  year={2002},
  publisher={APS}
}

@article{tam1969drag,
  title={The drag on a cloud of spherical particles in low Reynolds number flow},
  author={Tam, Christopher KW},
  journal={Journal of Fluid Mechanics},
  volume={38},
  number={3},
  pages={537--546},
  year={1969},
  publisher={Cambridge University Press}
}

@article{lundgren1972slow,
  title={Slow flow through stationary random beds and suspensions of spheres},
  author={Lundgren, To S},
  journal={Journal of fluid mechanics},
  volume={51},
  number={2},
  pages={273--299},
  year={1972},
  publisher={Cambridge University Press}
}

@article{vf1,
  title={Yield stress fluid flows: A review of experimental data},
  author={Coussot, Ph},
  journal={Journal of Non-Newtonian Fluid Mechanics},
  volume={211},
  pages={31--49},
  year={2014},
  publisher={Elsevier}
}

@article{vf2,
  title={High--Reynolds number wall turbulence},
  author={Smits, Alexander J and McKeon, Beverley J and Marusic, Ivan},
  journal={Annual Review of Fluid Mechanics},
  volume={43},
  number={1},
  pages={353--375},
  year={2011},
  publisher={Annual Reviews}
}

@article{vf3,
  title={Transition in wall-bounded flows},
  author={Lee, CB and Wu, JZ},
  journal={Applied Mechanics Reviews},
  volume={61},
  number={3},
  pages={030802},
  year={2008}
}
    \bibliographystyle{unsrt}		
			
	\end{document}